\documentclass[12pt,apj]{emulateapj}
\usepackage{apjfonts}

\usepackage{amsmath}

\newcommand{\kms}{\,{\rm km}\,{\rm s}^{-1}}	
\newcommand{\lya}{Ly$\alpha$}			
\newcommand{\hkpc}{\,h^{-1}\,{\rm kpc}}
\newcommand{\hmpc}{\,h^{-1}\,{\rm Mpc}}
\newcommand{\hGpc}{\,h^{-1}\,{\rm Gpc}}
\newcommand{\hgpc}{\,h^{-1}\,{\rm Gpc}}
\newcommand{\hubunits}{\kms\,{\rm Mpc}^{-1}}
\newcommand{\drp}{\Delta r_p}
\newcommand{\drz}{\Delta r_z}

\newcommand{\lyaf}{LyaF}
\newcommand{\lymas}{LyMAS}
\newcommand\per{{\tt Per}}
\newcommand\lper{{\tt per}}
\newcommand{\jcap}{JCAP}

\slugcomment{Submitted to the Astrophysical Journal}

\shorttitle{\lya\ Mass Association Scheme}
\shortauthors{Peirani et al.}

\begin{document}


\title{LyMAS: Predicting Large-Scale \lya\ Forest Statistics
       from the Dark Matter Density Field}


\author{S\'ebastien Peirani\altaffilmark{1},
 David H. Weinberg\altaffilmark{2,3},
St\'ephane Colombi\altaffilmark{1},
J\'er\'emy Blaizot\altaffilmark{4}, Yohan Dubois\altaffilmark{1}, Christophe Pichon\altaffilmark{1}
}

\altaffiltext{1}{Institut d'Astrophysique de Paris (UMR 7095: CNRS \& UPMC), 98 bis Bd Arago, 75014 Paris, France}
\altaffiltext{2}{Department of Astronomy, The Ohio State University, Columbus, OH}
\altaffiltext{3}{Sorbonne Universit\'es, Institut Lagrange de Paris (ILP), 98 bis Bd Arago, 75014 Paris, France}
\altaffiltext{4}{Universit\'e de Lyon, Lyon, F-69003, France; 
Universit\'e Lyon 1, Observatoire de Lyon, 9 avenue Charles Andr\'e, Saint-Genis Laval,
F-69230, France; CNRS, UMR 5574, Centre de Recherche Astrophysique de Lyon.}




\begin{abstract}
We describe \lymas\ (\lya\ Mass Association Scheme), a method of predicting 
clustering statistics in the \lya\ forest on large scales from moderate 
resolution simulations of the dark matter distribution, with calibration from 
high-resolution hydrodynamic simulations of smaller volumes. We use the 
``Horizon MareNostrum'' simulation, a $50\hmpc$ comoving volume evolved with the
adaptive mesh hydrodynamic code RAMSES, to compute the conditional probability
distribution $P(F_s|\delta_s)$ of the transmitted flux $F_s$, smoothed 
(1-dimensionally) over the spectral resolution scale, on the dark matter 
density contrast $\delta_s$, smoothed (3-dimensionally) over a similar scale. In
this study we adopt the spectral resolution of the SDSS-III Baryon Oscillation
Spectroscopic Survey (BOSS) at $z=2.5$, and we find optimal results for a dark
matter smoothing length $\sigma = 0.3\hmpc$ (comoving).  In its simplest form, 
\lymas\ draws randomly from the hydro-calibrated $P(F_s|\delta_s)$ to convert
dark matter skewers into \lya\ forest pseudo-spectra, which are then used to
compute cross-sightline flux statistics. In extended form, \lymas\ exactly
reproduces both the 1-dimensional power spectrum and 1-point flux distribution
of the hydro simulation spectra. Applied to the MareNostrum dark matter field,
\lymas\ accurately predicts the 2-point conditional flux distribution and flux 
correlation function of the full hydro simulation for transverse
sightline separations as small as $1\hmpc$, including redshift-space distortion
effects. It is substantially more accurate than a deterministic density-flux
mapping (``Fluctuating Gunn-Peterson Approximation''), often used for
large volume simulations of the forest. With the MareNostrum calibration, we
apply \lymas\ to $1024^3$ N-body simulations of a $300\hmpc$ and $1.0\hgpc$
cube to produce large, publicly available catalogs of 
 mock BOSS spectra\footnote{\parbox[t]{5.5cm}{The \lymas\ website can be found at:\\ http://www2.iap.fr/users/peirani/lymas/lymas.htm}\hfill \parbox[b]{1.6cm}{\resizebox{1.6cm}{!}{\includegraphics{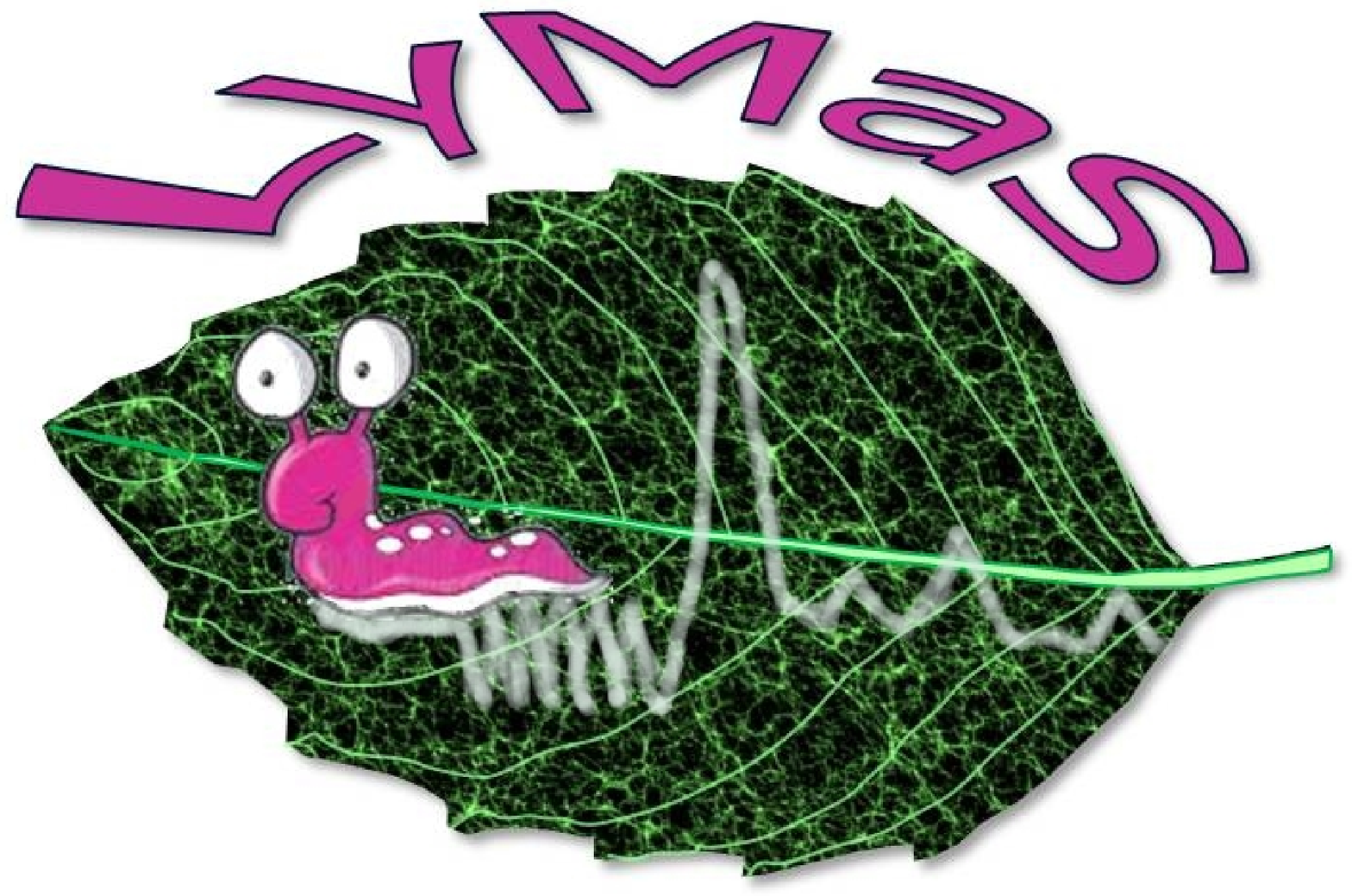}}\vspace*{-0.6cm}}}
that probe a large comoving volume.  LyMAS will be a powerful tool for interpreting 
3-d \lya\ forest data, thereby transforming measurements from BOSS and other 
massive quasar absorption surveys into constraints on dark energy, dark matter, 
space geometry, and IGM physics.
\end{abstract}





\section{Introduction}
\label{sec:intro}

The \lya\ forest -- fluctuating absorption produced by intergalactic
neutral hydrogen atoms in the spectra of background quasars --
is the most powerful probe of cosmic structure at redshifts
$z \approx 2-5$.  The physics of \lya\ forest absorption is
relatively well understood, thanks to numerical simulations
(e.g., \citealt{cen94,zhang95,hernquist96,miralda96} and related analytic
methods (e.g., \citealt{bi97,hui97a}) that accurately predict
its statistical properties with minimal assumptions beyond those
of the standard cosmological model.  The power spectrum, correlation
function, and probability distribution function (PDF) of 
\lya\ forest flux have been used to constrain the underlying matter
power spectrum and the thermal state of the intergalactic medium
(IGM), with corresponding constraints on cosmological parameters
and dark matter properties (e.g., 
\citealt{croft98,croft99,croft02,mcdonald00,mcdonald05,
zaldarriaga01,viel04,seljak05}).
These analyses have treated each sightline in isolation, as a collection
of individual 1-dimensional tracks through the intergalactic
gas distribution.  The Baryon Oscillation Spectroscopic Survey (BOSS)
of SDSS-III \citep{eisenstein11,dawson12} is opening a new ``3-dimensional''
frontier in \lya\ forest cosmology by measuring absorption on 
a grid of sightlines (150,000 quasars over 10,000 square degrees)
that is large enough and dense enough to enable accurate measurements
of transverse correlations across sightlines separated by tens
of $\hmpc$\footnote{Throughout the paper, we quote comoving
distances in $\hmpc$, where $h \equiv H_0/100\hubunits$.}.
BOSS and potential successor surveys
(e.g., BigBOSS, \citealt{schlegel11}) present a new theoretical challenge.
For instance, the recent first measurements of 
Baryon Acoustic Oscillations (BAO) in the  \lya\
forest fluctuations in the BOSS Data Release 9 \citep{busca12,slosar13} have
opened new perspectives in  constraining cosmological models,
 but the analysis and interpretation of such data require the development of efficient 
theoretical tools. In particular, to accurately predict the statistics of the \lya\ forest at large
scales for specified cosmological parameters, one must model
physics on the pressure-support scale (a.k.a. the ``Jeans scale'') of
the IGM  --- about $0.1-0.3\hmpc$ for typical values and roughly $0.25\hmpc$ at overdensity of 10 
(comoving at $z = 2-3$, see \citealt{peeples10}) --- 
while simulating volumes many hundreds
of $\hmpc$ on a side.  This paper proposes and tests a new method
of meeting this challenge.

Most \lya\ forest absorption arises in diffuse, highly photoionized
gas, which is heated by photoionization and cooled adiabatically
by expansion of the universe, leading to a tight correlation
between temperature and density \citep{katz96,hui97}.
This correlation enables the Fluctuating Gunn-Peterson Approximation
(FGPA; \citealt{weinberg98,croft98}), where the \lya\ optical
depth is tied to the local dark matter overdensity assuming
photoionization equilibrium and the \citet{gunn65} formula
for neutral hydrogen absorption.  The FGPA can be applied to
a log-normal density field created from Gaussian initial conditions
\citep{gnedin96,bi97} or to the dark matter density field of an 
N-body simulation,
with pressure effects incorporated by smoothing the field at the gas Jeans
scale or by using an approximate technique \citep{gnedin98} to incorporate
pressure into the N-body evolution.  Most efforts to systematically
model large scale \lya\ forest measurements and their dependence
on cosmological parameters have used some variant of this approach,
with full hydrodynamic simulations used to calibrate and test it
(e.g., \citealt{croft98,croft99,croft02,zaldarriaga01,mcdonald05}).
Growing computer power enables more ambitious approaches,
and \cite{viel06,viel10} used grids of hydrodynamic simulations
to model \lya\ forest statistics, but even these simulations
($60\hmpc$ on a side) are tiny compared to the volume probed by BOSS.

There are several reasons that the N-body or log-normal approaches
are unsatisfactory as a tool for modeling the 3-d \lya\ forest.
First, even without hydrodynamics, it is not currently feasible to
model Gpc$^3$ volumes while retaining good resolution on the gas
Jeans scale.  For example, even the heroic 3000$^3$-particle
simulations analyzed by \cite{slosar09} have an initial particle
spacing of $0.5\hmpc$, while smoothed particle hydrodynamic (SPH)
simulations require initial separations of $\sim 0.05\hmpc$ to
yield fully converged results \citep{lidz10,peeples10}.
Second, the choice of smoothing scale for the dark matter
inevitably produces some ambiguity in the predictions
\citep{zaldarriaga01}; \lya\ forest statistics are insensitive
to this choice, but not perfectly so \citep{viel02,peeples10}.
Third, the FGPA assumes a deterministic relation between
dark matter overdensity and continuum-normalized \lya\ flux
($F = e^{-\tau}$, hereafter simply ``flux''),
of the form $F = \exp\left[-A(\rho/\bar{\rho})^{2-0.6(\gamma-1)}\right]$,
where $(\gamma-1)$ is the index of the gas temperature-density
relation and $A$ is a normalization constant that depends on
several physical parameters (see \citealt{croft98,weinberg98,peeples10}).
In practice, there is scatter around this relation because
of peculiar velocities, shock heating of the gas above the
temperature-density relation, and pressure-induced differences
between the gas and dark matter distributions.
Peculiar velocities, the largest source of scatter
\citep{croft97,weinberg99,viel02}, can be incorporated into dark matter-based
approaches, but the latter two effects require a hydrodynamic
treatment.  Finally, even if the FGPA were a perfect approximation
on some scale (e.g., the gas Jeans scale), smoothing does not
commute with the non-linear transformation between density and flux,
so one cannot simply apply it to a lower resolution dark matter
simulation and recover a lower resolution version of the 
transmitted \lya\ flux.

We expect that the FGPA and log-normal methods, and even linear
perturbation theory with effective bias factors calibrated
on hydrodynamic simulations \citep{mcdonald03} are adequate
on sufficiently large scales, e.g., for predicting the baryon
acoustic oscillation (BAO) feature at $100\hmpc$ 
\citep{white03,mcdonald07,slosar09,white10, legoff10, greig11}, which is the
primary target of BOSS.  
\cite{font_ribera12} have developed an efficient semi-analytic method 
for generating mock data sets for the large-scale \lya\ forest, tailored
to reproduce the observed power spectrum and PDF,
which allows tests for systematics in analysis methods and calculation
of statistical errors and covariance matrices.
But BOSS and other 3-d \lya\ forest
surveys will also enable precise measurements of flux correlations
at scales of a few $\hmpc$ to a few tens of $\hmpc$, where
these approximations may not be accurate relative to the
measurement precision.  Large simulation volumes are still
needed in this regime, both to eliminate systematic effects of
the finite box size and to ensure that statistics are limited
by the available data not by the available simulation volume.
Deriving constraints and error bars requires making predictions
as a function of cosmological and IGM parameters.  
\cite{angulo10} have proposed an elegant and accurate method
for remapping non-linear dark matter distributions from 
one cosmology to another, so these parameter explorations
can be kept inexpensive if one can carry them out with
moderate resolution N-body simulations.

In this paper, we make explicit the separation-of-scale philosophy
that is already implicit in these alternative methods to predict
large-scale \lya\ forest statistics.  We concentrate on
the conditional flux probability distribution (PDF),
the probability $P(F_2 | F_1,\drp,\drz)$ of finding flux $F_2 = e^{-\tau_2}$
in a spectral pixel at location 2 given that the flux has the
value $F_1$ in a pixel with transverse (projected)
separation $\drp$ (see \citealt{miralda97}). 
Other two-point statistics such as the
flux correlation function 
$\langle (F_2-\bar{F})(F_1-\bar{F})\rangle$ can be computed
by integrating over this distribution and the unconditional
PDF (see \S\ref{sec:cpdf}).  It seems reasonable in
principle that to predict this distribution on large scales we should 
only need to resolve the matter distribution on scales
$r_{\rm res} \ll [(\drp)^2+(\drz)^2]^{1/2}$.  High-resolution
hydrodynamic simulations are needed to compute the relation
between the flux $F_s$, smoothed (1-dimensionally) on the
scale of instrumental resolution or some deliberately chosen
larger scale, and the matter density contrast $\delta_s$, smoothed
(3-dimensionally) on a scale larger than the simulation resolution
but smaller than the separations being probed.  
For the reasons discussed above, this relation
is not deterministic but stochastic, described by a
conditional probability distribution $P(F_s|\delta_s)$.  

Our conjecture is that one can predict $P(F_{s,2} | F_{s,1},\drp,\drz)$
for a given cosmological and IGM model by using a moderate resolution
N-body simulation to compute $P(\delta_{s,2} | \delta_{s,1},\drp,\drz)$
and convolving this with the local conditional distribution 
computed from high-resolution hydro simulations, $P(F_s|\delta_s)$
evaluated for $\delta_{s,1}$ and $\delta_{s,2}$.  Physically,
we are assuming that the fluxes at positions 1 and 2 are correlated
because the matter densities at these positions are correlated,
and that the {\it deviation} $F_{s,2} - \langle F_s|\delta_{s,2} \rangle$
from the conditional mean is uncorrelated with the {\it deviation}
$F_{s,1} - \langle F_s|\delta_{s,1} \rangle$.  The assumption
of uncorrelated deviations will certainly fail for closely 
neighboring pixels, whose fluxes will lie above or below the
conditional mean for the same reason, but we expect it to become
accurate quickly at larger separations.

We test this conjecture explicitly using a large, adaptive
mesh refinement (AMR) simulation of a $50\hmpc$ comoving
volume, realized with $1024^3$ dark matter particles and a
$1024^3$ top-level hydro grid, evolved with the AMR hydro code
RAMSES \citep{teyssier02}.  We use this simulation to calibrate
the $P(F_s|\delta_s)$ relation for several choices of dark matter
and \lya\ smoothing scales.
We then compare the \lya\ forest statistics computed
from the gas distribution of the simulation to those computed
from the dark matter distribution by the procedure described
above, which we refer to as LyMAS for 
\lya\ Mass Association Scheme\footnote{French readers may
appreciate the resonance with MoLUSC \citep{sousbie08},
which utilizes a similar style of dual-level calculation
to predict large scale galaxy clustering.}.
We also compare results to the deterministic FGPA approach
applied on the same smoothing scales,
finding that \lymas\ is significantly more accurate at
small separations.  
For compactness, we will sometimes abbreviate \lya\ forest
to \lyaf\ in the remainder of the paper.

In the next section we describe the hydro simulation and our
methods for creating \lyaf\ spectra and for extracting skewers
of the smoothed dark matter density field.  
Section~\ref{sec:cpdf} presents \lymas\ in its simplest form,
which is designed to predict flux statistics across sightlines
but does not produce realistic artificial spectra, because
it ignores correlated deviations of $F_s-\langle F_s|\delta_s\rangle$
along an individual line of sight.
In \S\ref{sec:pseudospectra} we describe an extended 
scheme that produces artificial spectra with a 1-d flux power
spectrum and 1-point PDF that (almost) exactly match those of the
full spectra from the hydro simulation while retaining
the accurate performance of \lymas\ for cross-sightline statistics.
In \S\ref{sec:appli} we describe the application of \lymas\ to
N-body simulations of a $300\hmpc$ and $1.0\hgpc$ comoving volume,
illustrating the impact of finite box size on flux correlation statistics.
We summarize our results and look towards future applications
in \S\ref{sec:discussion}.
Finally, in the website dedicated to LyMAS, we describe several sets of electronic tables
being made available with this paper, presenting the conditional
flux PDFs computed from the MareNostrum hydro simulation and 
the mock spectra extracted from the large N-body simulations.

\section{Hydrodynamic simulation and creation of spectra}
\label{sec:simulation}

In this section, we introduce the ``Horizon-MareNostrum''
 simulation, a high resolution cosmological N-body and hydrodynamics simulation
we have analyzed to generate our calibrations. We will then describe
our procedures to extract from this simulation the LyaF spectra as well as DM density fields.

\subsection{ The ``Horizon-MareNostrum'' simulation}

\begin{figure}
\begin{center}
\rotatebox{0}{\includegraphics[width=\columnwidth]{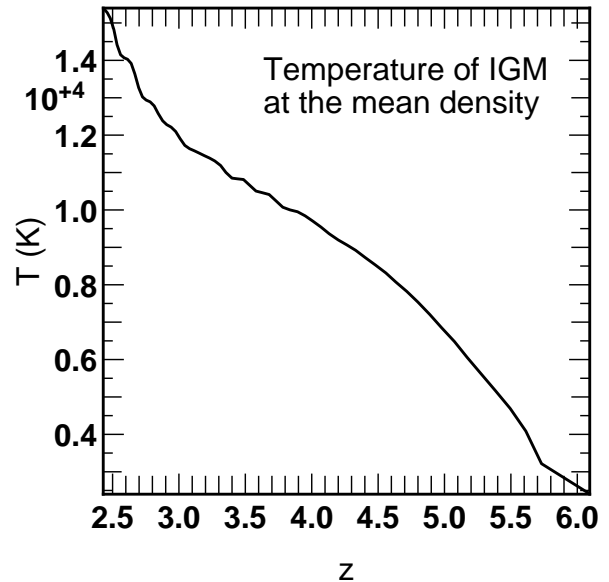}}
\caption{ Evolution of the mean temperature of the IGM at the mean density from redshift 6.5 to 2.5
in the MareNostrum simulation. The mean temperature of
the IGM tends to increase as a function of time mainly due to the global action of UV background, supernovae
feedback and shock-heating of the gas during galaxy formation processes.
}
\label{fig:MN0}
\end{center}
 \end{figure}

\begin{figure}
\begin{center}
\rotatebox{0}{\includegraphics[width=\columnwidth]{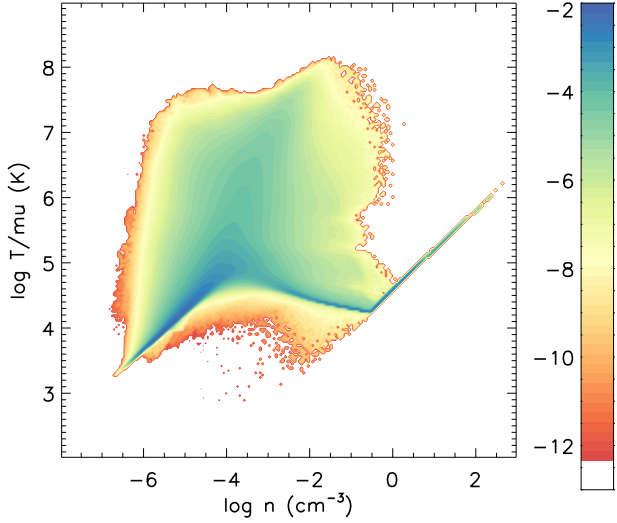}}
\caption{ Mass-weighted Temperature-density diagram of the MareNostrum simulation at $z=2.51$.
The mean density of the baryons is at $\bar n\simeq 10^{-5} \, \rm cm^{-3}$ at  this redshift.
}
\label{fig:MN}
\end{center}
 \end{figure}

The  Horizon-MareNostrum simulation (which we hereafter refer to
simply as MareNostrum) was run for a $\Lambda$CDM universe
using WMAP1 parameters \citep{spergel03}, namely
$\Omega_M=0.3$,  $\Omega_{\Lambda}=0.7$, $\Omega_B=0.045$,  $H_0=70.0$
km/s/Mpc, $n=1$ and $\sigma_8=0.9$. 
The simulation was performed in a periodic box of side $50\,h^{-1}$ Mpc
with $1024^3$ dark matter particles (e.g. with mass resolution
of $\sim 1.41 \times 10^7\,M_\odot$) and  an adaptive mesh
refinement (AMR) technique \citep{teyssier02} 
was used to keep the spatial resolution
fixed at $\sim 1 h^{-1}$ kpc in physical units.
The evolution of the thermodynamic properties of the gas depends on
a wide range of physical processes. In particular, MareNostrum 
incorporates metal-dependent cooling and 
photoionization heating from a UV background computed
via the prescriptions of \cite{haardt96}, starting at a
reionization redshift $z=8.5$.
A model of supernova feedback and metal enrichment has also been incorporated 
using the implementation described by \cite{dubois08}. 
The high-redshift \lya\ forest arises mainly in gas with overdensity
$1+\delta \la 50$, whose thermal evolution is governed primarily by
photoionization heating and adiabatic cooling from the expansion 
of the universe.  For reference, Figure~\ref{fig:MN0} shows the evolution
of the mean temperature of the IGM at the mean density 
(from $z=6.5$ to $z=2.5$), and Figure~\ref{fig:MN} shows the
mass-weighted temperature-density diagram of the MareNostrum
simulation at our analysis redshift $z=2.51$.
At densities $n \leq 10^{-4}\,{\rm cm}^{-3}$ one sees the usual
sharp locus of photoionized gas, with a small fraction of
gas shock heated above this relation.
At higher densities one sees shock-heated gas in halos
and the locus of radiatively cooled and star-forming gas in galaxies.
For further details of the MN simulation,
we refer the readers to previous papers that have 
used it for other analyses \citep{ocvirk08, devriendt10}.

\subsection{ Extracting \lyaf\ spectra}

We compute \lya{} absorption spectra along about one million lines of sight
cast through the $z = 2.51$ output of the MareNostrum simulation. At this 
redshift, the simulated volume allows us to probe \lya{} absorption lines 
over a range of $\sim 75$\AA{} around the observed wavelength $\lambda \sim 4268$\AA{}.
For a given sight-line, the opacity at observer-frame frequency $\nu_{obs}$
is $\tau(\nu_{obs}) = \sum_{\rm cells} n_{\rm HI} \sigma(\nu_{obs}) dl$, where the sum 
extends over all (leaf) cells traversed by the ray\footnote{In practice, we actually
extend this sum further by replicating the periodic simulated volume at both ends
of the rays.}, $n_{\rm HI}$ is the numerical density of neutral H atoms in each cell,
$\sigma(\nu_{obs})$ is the cross section of Hydrogen to \lya{} photons, and $dl$
is the physical cell size. We compute $\sigma(\nu_{obs})$ as:
\begin{eqnarray}
\sigma(\nu_{obs}) = f_{12}\frac{\pi e^2}{m_e c}  \times \frac{H(a,x)}{\sqrt{\pi} \Delta\nu_D},
\end{eqnarray}
where $f_{12}=0.4162$ is the \lya{} oscillator strength, $e$ and $m_e$ are the
 electron's charge and mass, $\Delta\nu_D = (2k_BT/m_H)^{1/2} \times \nu_\alpha /c$ is
 the Doppler frequency width due to thermal motions of atoms at temperature $T$,
 and $a = \Delta\nu_L / (2\Delta\nu_D)$ is the ratio of the natural line width
 ($\Delta\nu_L \sim 9.9\ 10^7 s^{-1}$) to the Doppler frequency width. The cross
 section is a function of frequency in the rest-frame of the absorbing gas
 cell $\nu_{cell} = \nu_{obs} (1+z)(1+v/c)$ (where $v$ is the cell's line-of-sight
 velocity), which we use to define the dimensionless frequency offset
 $x=(\nu_{cell}-\nu_\alpha) / \Delta\nu_D$. Finally, the Hjerting function $H(a,x)$ is:
\begin{eqnarray}
H(a,x) = \frac{a}{\pi} \int_{-\infty}^\infty \frac{e^{-y^2}dy}{a^2 + (x-y)^2}.
\end{eqnarray}

In practice, for each sight-line, we compute the opacity due to all cells
 on a high-resolution grid of observer-frame wavelength (with $d\lambda = 0.02$\AA).
 We then compute a high-resolution normalized flux $F(\lambda) = \exp(-\tau)$, which
 we integrate down to the targeted 1024 $\lambda$-bins.
At the observed frame wavelength $\lambda_{\rm Lya}(1+z) = 4268$\AA,
the resolution of the BOSS spectrographs \citep{eisenstein11,smee12}
is $\lambda/\Delta\lambda \approx 1740$, where $\Delta\lambda$ is
the FWHM of the line-spread function, implying $\Delta\lambda = 2.46$\AA.
The corresponding Gaussian dispersion is 
$\sigma = \Delta\lambda/(2\sqrt{2\ln 2}) = 1.045$\AA, 
which translates to a comoving distance of $0.696\hmpc$.
To mimic the effects of BOSS spectral resolution, therefore,
we convolve our extracted \lyaf\ spectra with a 1-d Gaussian
of dispersion $\sigma=0.696\hmpc$.
 
Note also that when generating our grid of flux, the MareNostrum simulation 
leads to a mean flux decrement with a slightly too high value.
Following standard practice in \lya\ forest modeling,
we match the observed mean decrement
at $z\sim2.5$ \citep{faucher-giguere08} by changing
the UV background intensity in post-processing, increasing
its intensity by a factor of about 1.3. 
The results should be virtually identical to those of a simulation
run with this higher intensity background during evolution,
since at IGM densities the photoionization heating rate is insensitive
to the UV background intensity \citep{thoul96,katz96}.

\subsection{Extracting dark matter skewers}

To extract dark matter skewers, we follow a 3-step process:

\begin{itemize}
\item Adaptive interpolation of the DM particle distribution on a high 
resolution grid (here a $1024^3$ grid for the Horizon-MareNostrum simulation
 and a $4096^3$ grid for the GADGET simulations of Section~\ref{sec:appli}).
 Note that this interpolation is performed after projection into
 redshift space if needed. Redshift space projection is performed along the $z$
 coordinate, so that we work in the distant observer limit.
\item Smoothing with a Gaussian window in Fourier space, using the
 fact that even when working in redshift space, periodicity of the
 box is preserved because we use the distant observer limit.
\item Extraction of the skewers from a grid of lines of sight
 aligned along the $z$ axis and positioned at random otherwise.
\end{itemize}
The adaptive interpolation method we use is described by
\cite{colombi07}, so we give here 
only a quick sketch of it. It is similar to SPH interpolation 
except that it is not a direct projection of the particles as 
an ensemble of round clouds on the grid. Indeed, the SPH 
interpolation method, if used as is, can behave poorly in
 underdense regions, where it can leave totally empty grid
 sites. To avoid this defect, each grid site is considered
 as a virtual particle and the $N_{\rm SPH}$ closest particles
 to it are found (here $N_{\rm SPH}=32$). Then we perform
 a calculation of the density at the center of the cell
 using a weight for each particle $i$ that is roughly
 proportional to $W(r_i)$, where $r_i$ is the distance 
of the particle to the center of the cell and where $W(r)$
 is a standard SPH kernel function \citep{monaghan92}. We
 use the term ``roughly'' because the actual interpolation
 scheme is more complex than that: (i) special treatment
 is performed when the number of particles in the cell,
 $N_{\rm p}$, exceeds $N_{\rm SPH}$: in this case, we perform 
the interpolation over the $N_{\rm p}$ particles instead of
 $N_{\rm SPH}$: this is to minimize the deviation from total
 mass conservation discussed below; (ii) to make sure that
 the total contribution of each particle equals its mass $m$,
 a renormalization of its local contribution to each cell is 
performed {\em a posteriori} (i.e., after calculating its
 contribution to each individual cell). Our scheme does
 not guarantee that all the particles contribute: there
 might be some cases where a particle could not contribute
 to any grid site, which means that the total mass is not
 conserved exactly. This effect is however small, usually
 of the order of $0.1$ percent (it is fixed by an overall
 renormalization of the grid density).

The adaptive interpolation scheme is relatively expensive, but
it preserves high-resolution information in overdense regions
of the dark matter field.  We have not tested \lymas\ with simpler
density assignment methods such as cloud-in-cell, so we caution 
that our favorable findings below might not apply in this case.

\begin{figure}
\begin{center}
\rotatebox{0}{\includegraphics[width=\columnwidth]{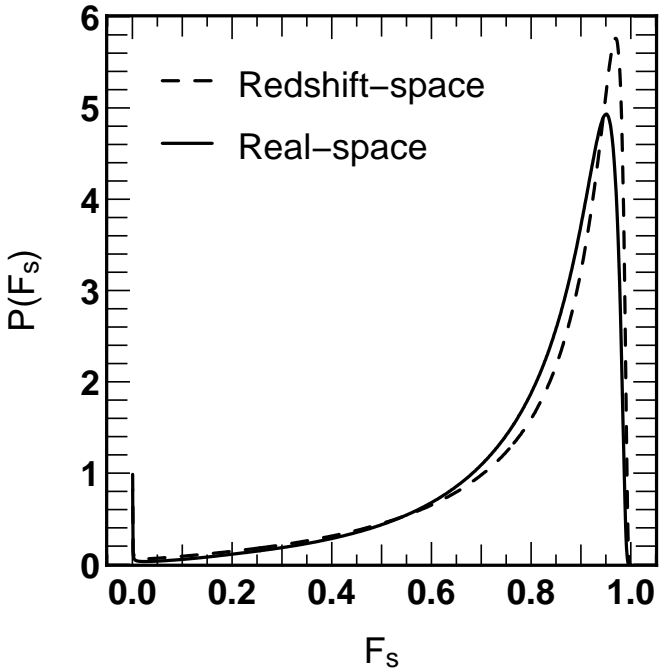}}
\caption{Unconditional PDF of the spectra extracted from
the hydro simulation at redshift $z=2.51$, in 
real space (solid) and redshift space (dashed).
The effective optical
depth $\tau_{\rm eff} = -\ln \bar{F_s}$ is 0.2301 and 0.2288
for real and redshift space, respectively, in good agreement
with the observational estimate of \cite{faucher-giguere08}
at $z=2.5$.
Spectra have been smoothed with a Gaussian of $\sigma = 0.696\hmpc$
(comoving) to mimic BOSS spectral resolution.
}
\label{fig:uncond_pdf}
\end{center}
 \end{figure}

\section{Predicting conditional flux distributions}
\label{sec:cpdf}

\begin{figure*}
\begin{center}
\rotatebox{0}{\includegraphics[width=16.5cm]{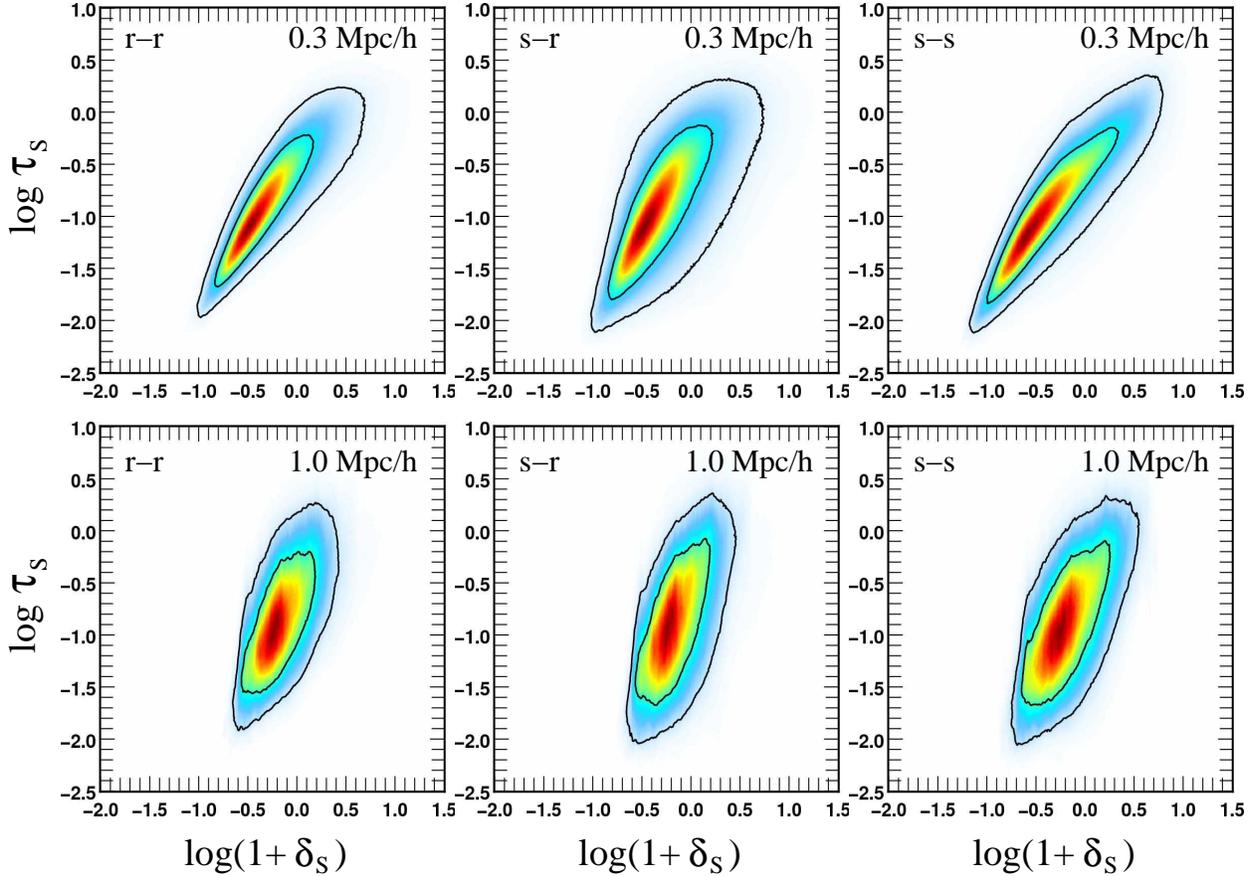}}
\caption{Correlation between optical depth $\tau_s=-\ln F_s$ in 
the hydro spectra and the smoothed DM overdensity $(1+\delta_s)$
at the same location along the line of sight, for comoving
DM smoothing lengths of $0.30\hmpc$ (top row) and $1.0\hmpc$
(bottom row).  Colors show the density of pixels in the
$\log \tau_s - \log(1+\delta_s)$ plane, and contour lines mark
areas enclosing 68.27\% 95.45\% respectively.
Left panels show the case with spectra and DM density both
computed in real space (peculiar velocities set to zero),
middle panels show redshift space spectra and real space
density, and right panels show both computed in redshift space.
}
\label{fig:scatter_plots}
\end{center}
 \end{figure*}

Our approach to predicting \lyaf\ flux statistics is based on
the conditional distribution of flux given density.
Figure~\ref{fig:uncond_pdf} plots the {\it unconditional} flux PDF
of the LyaF spectra extracted from the hydro simulation, in real
space and in redshift space, where the spectra have been smoothed
to BOSS resolution.
Figure~\ref{fig:scatter_plots} plots contours of the 
joint distribution of optical depth
$\tau_s = -\ln F_s$ and overdensity $(1+\delta_s)$, where
the subscript $s$ emphasizes that the flux is smoothed
(in one dimension) to the BOSS spectral resolution and the
density contrast is smoothed with a 3-d Gaussian of radius
$\sigma = 0.3\hmpc$  (top panels) or $1.0\hmpc$ (bottom
panels).  The top left panel shows a tight correlation
between the real space smoothed overdensity and the
``real space'' \lyaf\ flux, i.e., the flux computed with
all gas peculiar velocities (but not thermal broadening)
set to zero.  This tight correlation, first noted by
\cite{croft97}, provided the initial motivation for the FGPA.
Croft et al.\ noted that the scatter in the correlation goes
up when one includes peculiar velocities of the gas, 
which shift the line of sight location of absorption 
features and raise or lower optical depths through
convergence or divergence in velocity space.
We see this effect in the top middle panel of
Figure~\ref{fig:scatter_plots}.  However, in the right panels
we show the correlation when the dark matter distribution is
{\it also} computed in redshift space: the scatter decreases,
producing a correlation that is essentially just as tight
as the one obtained in real space.

The same trends hold for a $1.0\hmpc$ smoothing length (lower panels),
but in this case the range of densities is compressed, and
the correlation is steeper.  As a result, the scatter in 
optical depth has larger impact relative to the trend with density.
For our calculations below, we have experimented with a range
of smoothing lengths, and we find that the best results
(i.e., the most accurate predictions for flux statistics) are
obtained with a smoothing length in the range $\sim 0.2 - 0.35\hmpc$.
This is comparable to the Jeans scale for $1+\delta = 10$
\citep{peeples10} and somewhat smaller than the BOSS spectra
resolution scale of $\sigma = 0.696\hmpc$.
We have not tested whether smaller smoothing lengths would
give better results for higher resolution spectra.
However, we also carry out most of our tests for $1.0\hmpc$ smoothing,
as this still works quite well, and it can be applied to
N-body simulations with lower resolution.

\begin{figure*}
\begin{center}
\rotatebox{0}{\includegraphics[width=16cm]{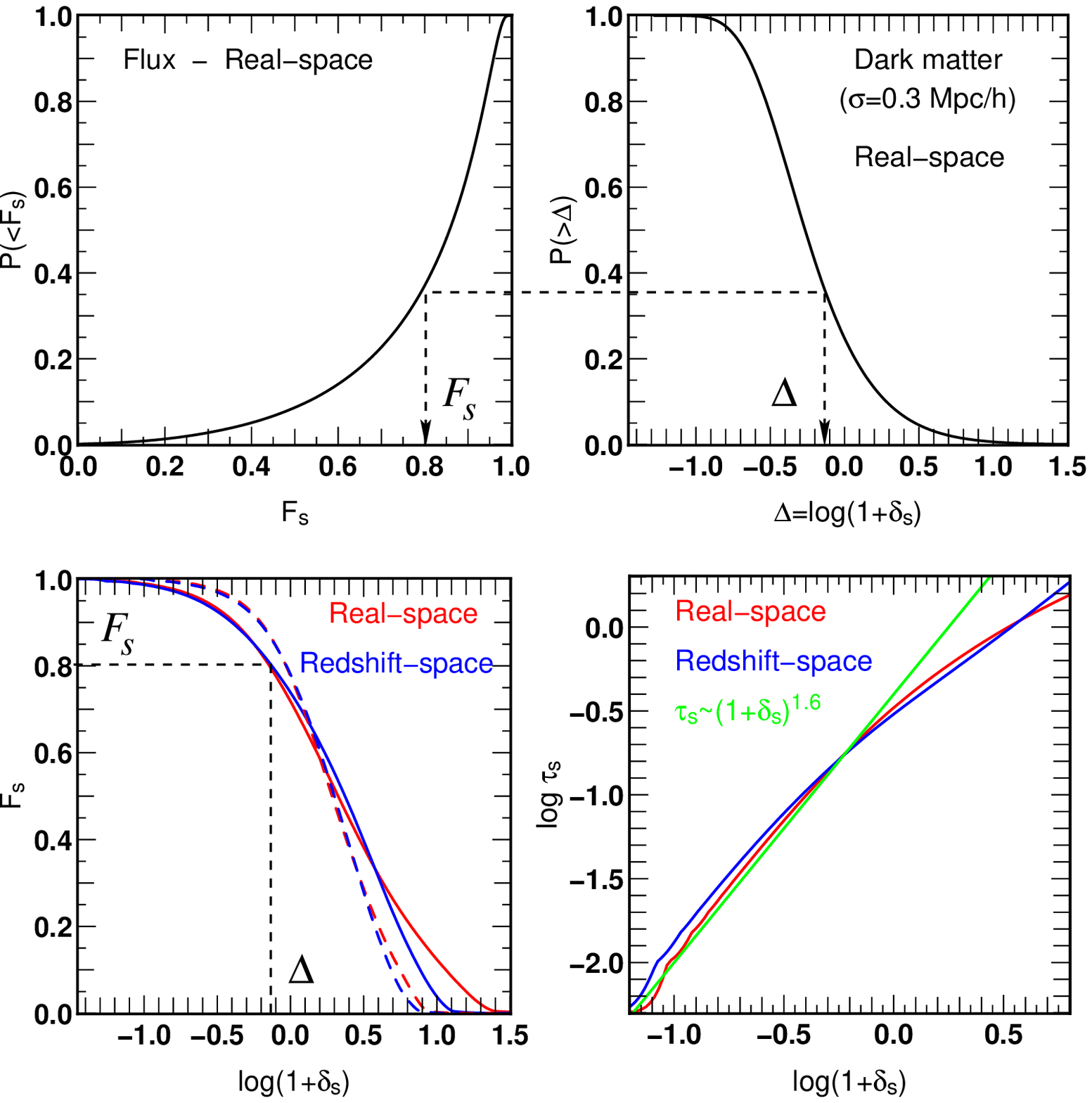}}
\caption{Summary of the procedure to get deterministic relations 
$F(\delta)$. Upper panels plot the cumulative distributions of
$F_s$ (left) and $\delta_s$ (right).  We match a flux $F_s$ to
the density contrast $\delta_s$ at the same location on the
cumulative distribution, with higher fluxes (lower absorption)
matched to lower densities.  The lower left panel shows the
resulting $F_s(\delta_s)$ relations computed in real and
redshift space (blue and red solid curves, respectively).
Solid lines show results for DM smoothing length $\sigma=0.3\hmpc$
and  dashed lines show results for $1.0\hmpc$.  In the 
lower right panel we plot $\tau_s = -\ln F_s$ against
overdensity $(1+\delta_s)$ for the $\sigma=0.3\hmpc$ case.
The green line shows the logarithmic slope predicted
by the FGPA.
}
\label{fig:determ_rel}
\end{center}
 \end{figure*}

The FGPA converts DM density to optical depth using a physical
model motivated by photoionization equilibrium assuming that all
gas contributing to the \lyaf\ lies on a temperature-density relation 
$T \propto (\rho/\bar{\rho})^{\gamma-1}$, leading to
$\tau \propto (\rho/\bar{\rho})^{2-0.7(\gamma-1)}.$
This relation makes sense for modeling high-resolution spectra,
but it does not necessarily apply at low resolution because
of the non-linear relation between flux and optical depth,
and even in the high-resolution case it omits some physical effects.
Alternatively, one can derive an ``optimal'' deterministic
relation between smoothed density and smoothed flux that
is calibrated on hydro simulations, a strategy advocated by 
\cite{Gallerani11}.  
A straightforward choice for an ``optimal'' mapping
is to find the unique monotonic
relation between the smoothed matter density and the smoothed
flux such that the spectra generated from the dark matter have the same
unconditional 1-point flux PDF as the spectra from the hydro
simulation.  We do so by matching the corresponding cumulative
distributions to a pixel with DM contrast $\delta_s$ we assign
a flux $F_s$ such that
\begin{equation}
\int_{0}^{F_s} P(F'_s)dF'_s = \int_{\delta_s}^\infty P(\delta'_s)d\delta'_s,
\label{eqn:deterministic}
\end{equation}
where $P(F_s)$ and $P(\delta_s)$ are the 1-point PDFs of flux
and DM density measured from the simulation.
Figure~\ref{fig:determ_rel} illustrates this procedure
graphically and shows the resulting mapping, which is slightly
different in real space and redshift space because the PDFs
are different in the two cases (see Fig.~\ref{fig:uncond_pdf}).
The mapping is also different if we consider the $1.0\hmpc$
DM smoothing rather than the $0.3\hmpc$ smoothing.
Well after reionization, the interplay between photoionization
heating and adiabatic cooling produces a temperature-density
relation $T \propto \rho^{0.6}$ in the diffuse photo-ionized
medium that dominates the \lyaf\ \citep{katz96,hui97},
and in this case (which applies in our simulation)
the FGPA predicts $\tau \propto \rho^{1.6}$.
Our deterministic relations for $0.3\hmpc$ smoothing
have this logarithmic slope at low density, 
but the relation flattens somewhat at higher density,
with lower optical depth than the $(1+\delta_s)^{1.6}$
extrapolation predicts.  This flattening may be partly
a consequence of working at BOSS resolution rather than
high resolution.
We will use the deterministic mappings plotted here as a
foil for evaluating the success of the \lymas\ probabilistic
scheme.

\begin{figure*}
\begin{center}
\rotatebox{0}{\includegraphics[width=16.5cm]{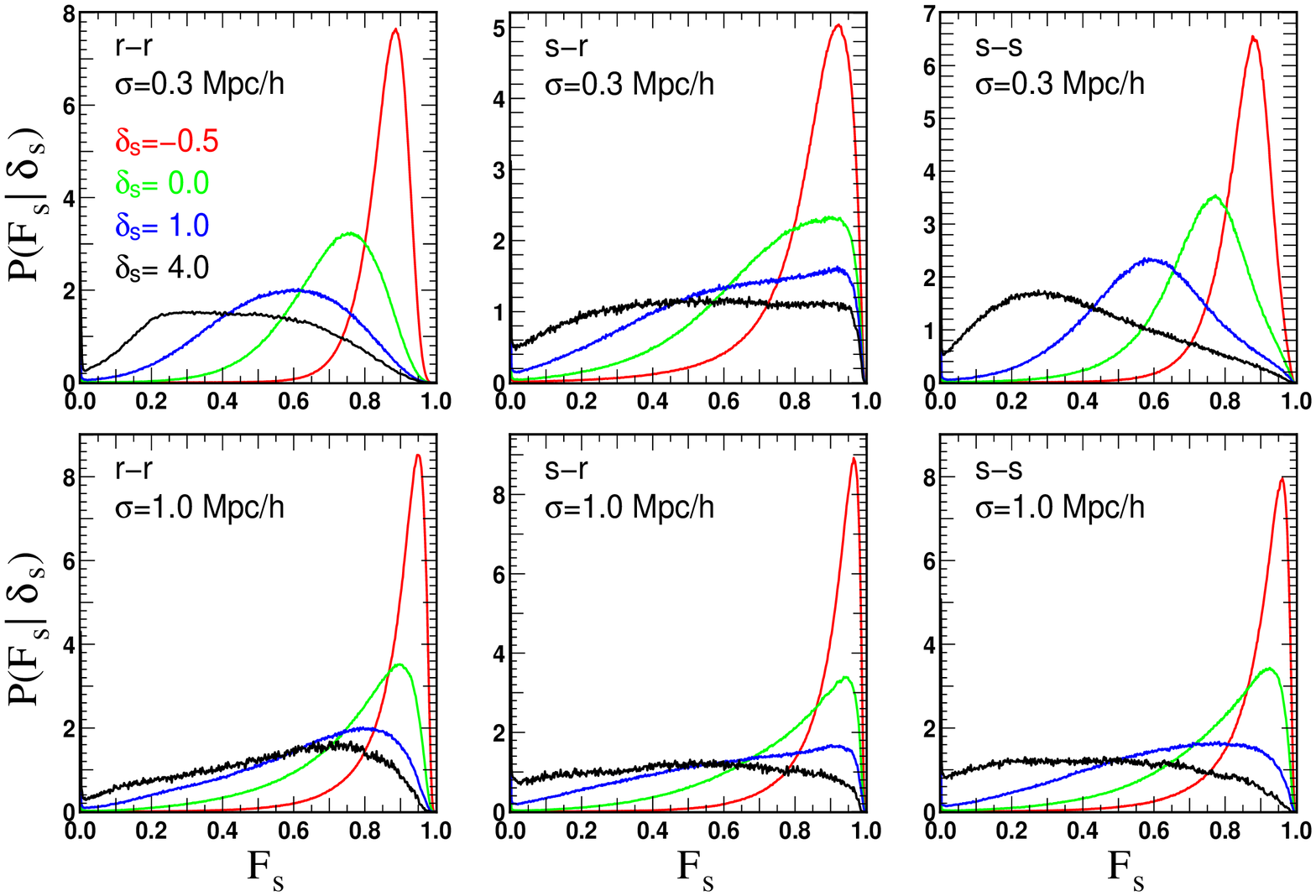}}
\caption{Conditional probability distributions $P(F_s|\delta_s)$
for the smoothed flux at different values of the density contrast
$\delta_s = -0.5$, 0.0, 1.0, and 4.0, as labeled.  
Upper and lower panels correspond to DM smoothing lengths
$\sigma = 0.3\hmpc$ and $1.0\hmpc$, respectively.
Left panels are in real space, right panels are in redshift space,
and middle panels use redshift-space spectra but real-space
DM densities.
}
\label{fig:prob_cond_dm}
\end{center}
 \end{figure*}

The scatter in Figure~\ref{fig:scatter_plots} shows that the
relation between DM density and flux is not, in fact, deterministic.
Figure~\ref{fig:prob_cond_dm} plots the conditional probability
distributions $P(F_s|\delta_s)$ for dark matter density contrasts
$\delta_s = -0.5$, 0, 1, and 4; each curve represents a vertical cut
through the corresponding panel of Figure~\ref{fig:scatter_plots}.
Beginning with the top row, for DM smoothing length $\sigma=0.3\hmpc$,
we see that these conditional distributions are quite broad,
though they always show the expected trend of lower average flux
for higher DM density.  The middle panel, based on redshift-space
spectra but real-space DM densities, has broader distributions as 
expected.  However, when we use redshift-space DM densities
(right panel), the distributions are if anything sharper and
more differentiated than those obtained for real-space spectra
and real-space DM densities (left panel).
The bottom row, for DM smoothing $\sigma=1.0\hmpc$, shows similar
trends, but the curves are less differentiated than those
for $\sigma=0.3\hmpc$.  The distributions for $\delta_s = -0.5$, 0, and 1
have maxima at nearly the same flux level ($F_s=0.8-0.95$ in the
right panel), and in redshift-space the $\delta_s=4$ curve is 
nearly flat over a wide range of flux, so its maximum is poorly defined.
This difference is not surprising, since the $\sigma=1.0\hmpc$ smoothing
scale is larger than the BOSS spectral resolution scale,
and smoothing the DM density to this scale therefore loses information
that is relevant to predicting the BOSS-resolution flux.
We will nonetheless find that the \lymas\ scheme with $1.0\hmpc$
DM smoothing outperforms the deterministic scheme with $0.3\hmpc$
smoothing.

As described already in the introduction, the philosophy of \lymas\
is to assume that the fluxes along separate lines of sight can
be drawn independently from the conditional probability
distributions $P(F_s|\delta_s)$, with the flux correlations
across lines of sight arising from correlations in the underlying
DM density field  
(coherent fluctuations in the UV background or the IGM
equation-of-state could also produce large scale flux correlations;
these effects can be incorporated into \lymas\ simulations, but
we have not done so here).
To compute 2-point flux statistics via \lymas, we create artificial
spectra along each of the DM skewers extracted from the simulation,
assigning to each pixel a flux that is randomly drawn from
the conditional distribution $P(F_s|\delta_s)$ that we have
computed from the hydro simulation and illustrated in 
Figure~\ref{fig:prob_cond_dm}.  The creation of artificial
spectra is not strictly necessary for computing these flux statistics;
it is effectively a convenient Monte Carlo method of doing the relevant
integrals over the probability distributions.
As discussed in \S\ref{sec:pseudospectra} below, the
artificial spectra created in this simple manner are unrealistic
because they ignore additional correlations (beyond those from
DM alone) that arise in neighboring pixels along a single line
of sight.  In \S\ref{sec:pseudospectra} we present a method for
creating realistic artificial spectra that have the 
correct degree of coherence along individual lines of sight,
but this more complex method is not needed for predicting
flux statistics in the idealized case, with no contribution from
observational effects such as noise and continuum fitting errors.

\begin{figure*}
\begin{center}
\rotatebox{0}{\includegraphics[width=16.5cm]{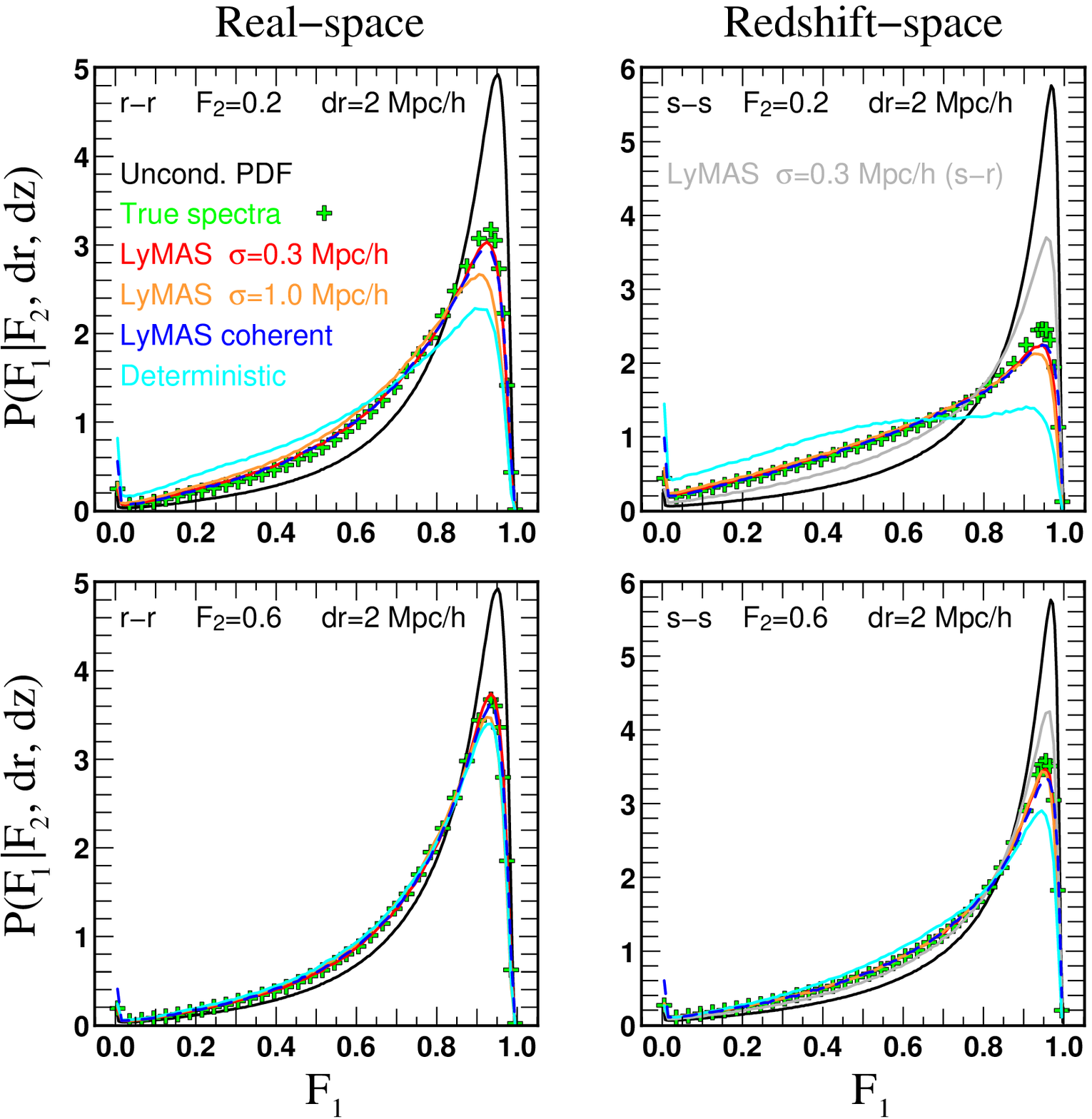}}
\caption{Two-point conditional flux PDF $P(F_1|F_2,dr,dz)$ computed for
$dr=2.0\hmpc$ and $dz=0$, computed in real space (left panels)
and redshift space (right panels).  Upper and lower panels show
conditioning flux $F_2 = 0.2\pm 0.005$ or $0.6\pm 0.005$, respectively.
In each panel we show the conditional PDF for the true hydro
spectra (green crosses), \lymas\ results for DM smoothing 
$\sigma = 0.3\hmpc$ (red) and $1.0\hmpc$ (orange), and results
for the deterministic method with $\sigma=0.3\hmpc$ (cyan).
Black curves show the unconditional PDF for comparison.
Blue dashed curves, which are nearly indistinguishable from
the red curves, show results from the ``coherent''
\lymas\ scheme described in \S\ref{sec:pseudospectra}.
In the right panels, grey curves show the result of using
the real-space DM density field to predict the redshift-space spectra.
}
\label{fig:prob_cond_r}
\end{center}
 \end{figure*}

Figure~\ref{fig:prob_cond_r} plots conditional PDFs for fluxes of
pixels at the identical redshift along lines of sight with transverse
comoving separation of $2\hmpc$: $P(F_1|F_2,dr,dz)$  is the probability
of finding flux $F_1$ given that the pixel at $dr=2\hmpc$, $dz=0$
has flux $F_2 = 0.2$ (top panels) or 0.6 (bottom panels).
In practice, we compute these distributions for pixels with
$F_2 = 0.2 \pm 0.005$ or $0.6 \pm 0.005$.
Left panels show the case with spectra and DM densities both
computed in real space (i.e., with all peculiar velocities
set to zero), and right panels show the case with both spectra
and density fields computed in redshift space.
In each panel, the solid black curve shows the unconditional
1-point PDF.  As expected, imposing the condition $F_2 = 0.2$,
i.e., strong absorption along the neighboring line of sight,
increases the probability of $F_1$ being low.  The peak of the 
conditional distribution is depressed in amplitude, but it is only
slightly shifted in location relative to the unconditional PDF.
Interestingly, the impact of the $F_2$ condition is stronger
in redshift space than in real space, presumably reflecting the
additional effects of coherence in the velocity field.
For $F_2=0.6$, which still represents fairly strong absorption
(about 15th-percentile in the cumulative distribution), we
see similar effects but reduced in magnitude.

Even for this fairly small transverse separation, 
the \lymas\ scheme with $\sigma=0.3\hmpc$ reproduces the
results from the true hydro spectra\footnote{We hereafter refer
 ``true hydro spectra'' or ``true spectra'' 
to as spectra derived from the MareNostrum hydro simulation.} almost perfectly,
with the most noticeable error being a small deviation
in the height of the peak.  The deterministic scheme
is considerably less accurate, especially in redshift
space and for the lower $F_2$.
Remarkably, despite the broader conditional distributions
seen in Figure~\ref{fig:prob_cond_dm}, \lymas\ with
$\sigma=1.0\hmpc$ is nearly as accurate as with $\sigma=0.3\hmpc$
when computing the conditional flux PDF at these separations, 
and it is again much more accurate than the deterministic relation.
The blue-dashed curve in all panels shows results for the
``coherent'' pseudo-spectrum method described in 
\S\ref{sec:pseudospectra}; for now, we simply note that
they are essentially identical to those from our ``minimal''
\lymas\ scheme with the same DM smoothing length.
Grey curves in the right panels show the effect of 
using the real-space DM density to predict the redshift-space
flux.  The performance is much worse than using the
redshift-space density field, as expected given our
previous results.  For all future plots, we will use
only the redshift-space hydro spectra and the redshift-space
DM density fields.

We note that all four of the schemes shown in 
Figure~\ref{fig:prob_cond_r} reproduce the unconditional
1-point PDF of the true spectra by construction, up to
tiny statistical fluctuations associated with the finite
number of spectra that we have created.

\begin{figure*}
\begin{center}
\rotatebox{0}{\includegraphics[width=14.5cm]{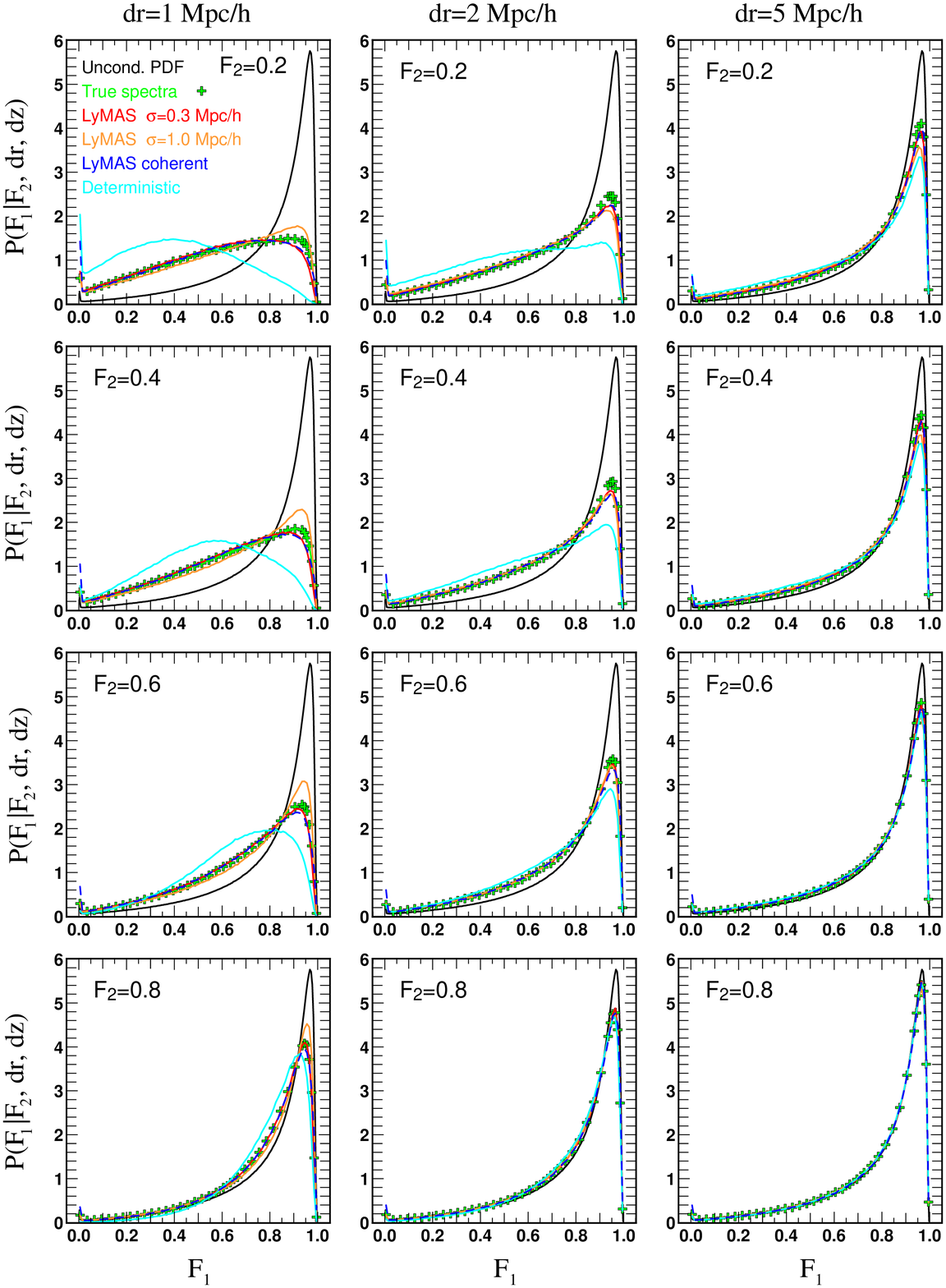}}
\caption{Two-point conditional flux PDF, plotted
as in Fig.~\ref{fig:prob_cond_r} but for a wider range of separations
and conditioning flux values, with all computations in redshift space.
Left, middle, and right columns show transverse separations
$dr=1.0$, 2.0, and $5.0\hmpc$, respectively, with $dz=0$ in all cases.
Rows from top to bottom show conditioning flux $F_2=0.2$, 0.4, 0.6, and 0.8.
}
\label{fig:prob_cond_s}
\end{center}
 \end{figure*}

Figure~\ref{fig:prob_cond_s} presents a broader view of the
the conditional flux distributions.  We now consider three
different transverse separations, $dr = 1$, 2, and $5\hmpc$,
and four different conditioning flux values, $F_2=0.2$, 0.4, 0.6, 
and 0.8.  Two of the panels in the middle column repeat the
two right-hand panels of Figure~\ref{fig:prob_cond_r}.
We can see a number of trends, all in qualitative agreement
with expectations.  The impact of the conditioning flux
is strongest when the transverse separation is small,
though it is still noticeable at $dr=5\hmpc$.
As the conditioning flux $F_2$ is increased, the 
conditional PDF of $F_1$ shifts towards higher
flux values.  For $F_2=0.8$ and $dr=5\hmpc$, the 
conditional PDF nearly matches the unconditional PDF.
\lymas\ predictions are more accurate than those from
the deterministic scheme in every case, though the 
difference becomes smaller as the conditional PDF
approaches the unconditional PDF.  For $\sigma=0.3\hmpc$,
\lymas\ is impressively accurate even for a transverse
separation $dr=1\hmpc$.  For $\sigma=1.0\hmpc$, the predictions
are noticeably less accurate at $dr=1\hmpc$ than at $dr=2\hmpc$,
though still much more accurate than those of the deterministic
scheme.  The coherent spectrum method of \S\ref{sec:pseudospectra}
gives results that
are essentially indistinguishable from those of the
minimal \lymas\ in all cases.

\begin{figure*}
\begin{center}
\rotatebox{0}{\includegraphics[width=14cm]{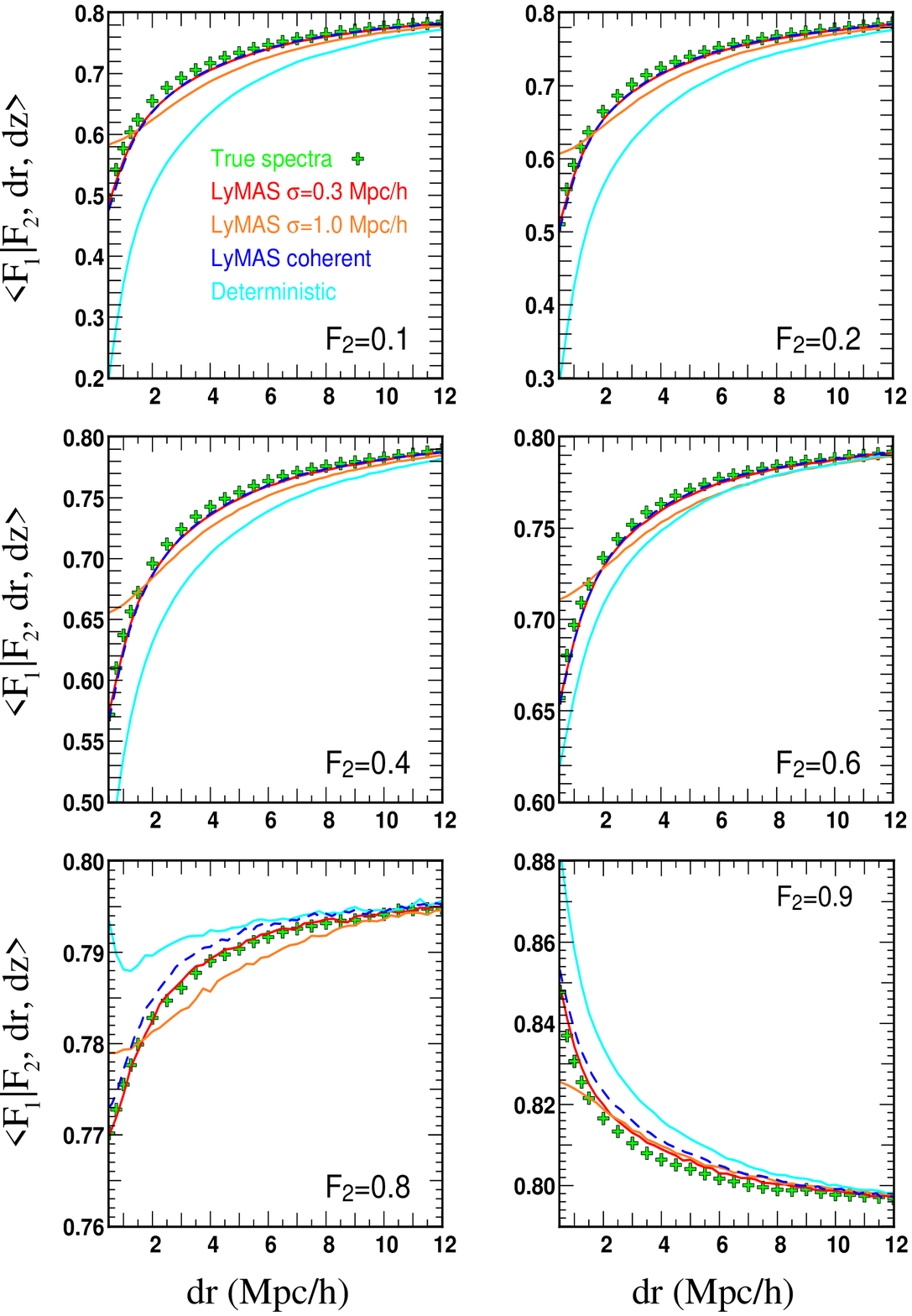}}
\caption{The conditional mean flux $<F_1|F_2,dr,dz>$ 
as a function of transverse separation $dr$ (with $dz=0$),
for six values of $F_2$ as labeled.  We again show results from 
the true hydro spectra as green crosses, with curves showing
results from \lymas\ with $\sigma=0.3\hmpc$ (red) or $1.0\hmpc$ (orange)
or the deterministic method with $\sigma=0.3\hmpc$ (cyan).
Blue dashed curves, virtually indistinguishable from the red curves,
show results from the coherent \lymas\ scheme described 
in \S\ref{sec:pseudospectra}.
All computations are in redshift space.
}
\label{fig:mean_cond_f}
\end{center}
 \end{figure*}

Conditional distributions like those in Figure~\ref{fig:prob_cond_s}
provide a comprehensive description of the two-point correlations
in the \lyaf\ flux field.
However, they are difficult to measure from spectra with significant
noise, since the underlying flux PDFs are convolved with the noise
distribution.  Figure~\ref{fig:mean_cond_f} shows a higher-level
summary statistic, the conditional mean flux 
$\langle F_1\rangle = \int_0^1 F'_1 P(F'_1|F_2,dr,dz)dF'_1$, plotted as
a function of transverse separation $dr$ for various choices of
the conditioning flux $F_2$.  The effects of noise
on the PDF of $F_1$ should average to zero when computing
the conditional mean, making this a more easily measurable
statistic (it also requires fewer lines of sight to give
a useful measurement).
However, a proper comparison to observations would have to account
for the effects of noise on the conditioning flux $F_2$, which
do {\it not} cancel by averaging; we ignore these effects here.
Note that the vertical scales are different in each panel.
As expected, the conditional mean flux approaches the unconditional
mean $\bar{F}=0.796$ at large $dr$, and it dips (or, in the
case of $F_2=0.9$, rises) towards the value of $F_2$ for small
transverse separation.  Starting from the value at $dr=0.5\hmpc$,
the conditional mean has typically recovered about halfway towards 
the unconditional value by $dr=3\hmpc$.
\lymas\ with $\sigma=0.3\hmpc$ reproduces the results from
the true hydro spectra accurately in all cases even down
to $dr=0.5\hmpc$.  It is substantially more accurate than
the deterministic scheme, which tends to strongly overestimate
the impact of the $F_2$ condition at small transverse separations.
With the finer discrimination allowed by this summary statistic,
we see that the $\sigma=1.0\hmpc$ smoothing length makes \lymas\
noticeably less accurate even at separations of $6-8\hmpc$,
though the largest discrepancy arises at $dr \leq 1\hmpc$,
and even with the larger smoothing length \lymas\ is almost
always more accurate than the deterministic scheme
computed with the optimal smoothing length.

\begin{figure*}
\begin{center}
\rotatebox{0}{\includegraphics[width=12.5cm]{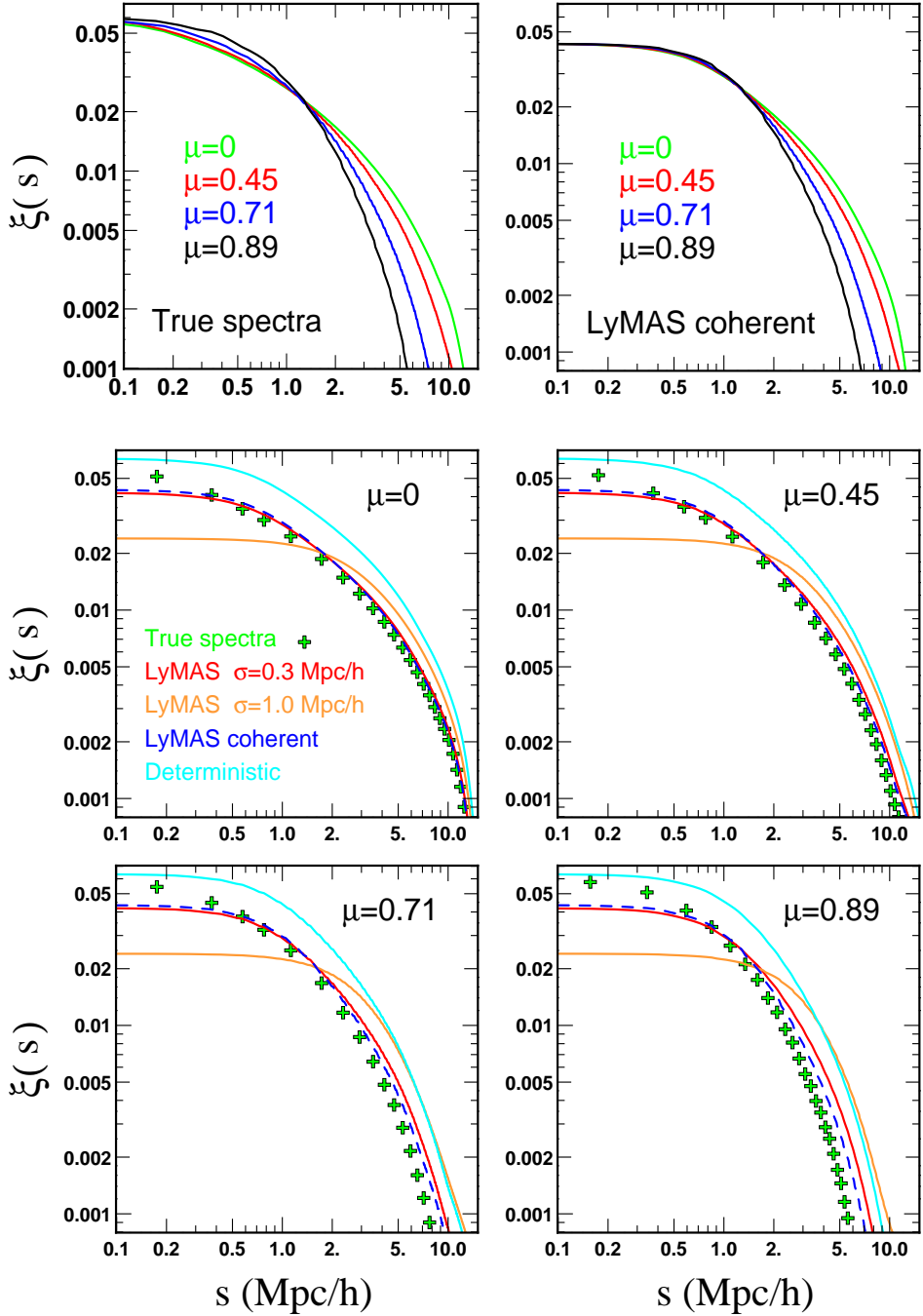}}
\caption{The correlation function 
$\xi=\langle F_1(r,z)F_2(r+dr,z+dz)\rangle /\langle F\rangle^2 -1$ 
as a function of the separation $s=(dr^2+dz^2)^{1/2}$,
for four angles: $dz=0$ ($\mu=0$),  $dr=0.5dz$ ($\mu\simeq 0.45$), 
$dr=dz$ ($\mu\simeq 0.71$)
and  $dr=2dz$ ($\mu\simeq 0.89$).
Again, we present results from 
the true hydro spectra as green crosses, with curves showing
results from \lymas\ with $\sigma=0.3\hmpc$ (red) or $1.0\hmpc$ (orange)
or the deterministic method with $\sigma=0.3\hmpc$ (cyan).
Blue dashed curves
show results from the coherent \lymas\ scheme described 
in \S\ref{sec:pseudospectra}.
All computations are in redshift space. 
Top panels show all four $\mu$ values on a single plot
for the true spectra (left) and coherent \lymas\ with
$\sigma = 0.3\hmpc$ (right).
 }
\label{fig:correlation_func}
\end{center}
 \end{figure*}

Finally, Figure~\ref{fig:correlation_func} shows the
variations in redshift space of the correlation function  
$\xi(s,\mu)=\langle F_1(r,z)F_2(r+dr,z+dz)\rangle/\langle F\rangle^2 -1$ 
as a function of the separation $s=(dr^2+dz^2)^{1/2}$ and angle
$\mu$ defined for a pair of pixels ($i$, $j$) by $\mu=(r_i-r_j)_{||}/r$, 
where $r=|r_i-r_j|$ and 
$(r_i-r_j)_{||}$  the component along the line of sight. 
In the present case, we 
have considered four distinct angles: 
$dz=0$ (i.e. $\mu=0$),  $dr=0.5dz$ ($\mu\simeq 0.45$),
 $dr=dz$ ($\mu\simeq 0.71$) and  $dr=2dz$ ($\mu\simeq 0.89$).
Again, we find that results obtained from the coherent \lymas\ 
scheme with $\sigma=0.3\hmpc$
are in good 
agreement with those from the true hydro spectra apart from very small
separations ($s\lesssim 0.5\hmpc$) where  \lymas\ tends to underestimate  
the correlation function computed from the true spectra.
As shown in the top panels, the $\mu$-dependence of $\xi(s,\mu)$ is quite
strong, the signature of redshift-space distortions in the \lya\ forest.
This effect is seen in the observational analysis of BOSS spectra
by \cite{slosar11}.  With $\sigma=0.3\hmpc$, \lymas\ predicts this
redshift-space distortion accurately, though it performs somewhat worse 
at high values of $\mu$.  With $\sigma=1.0\hmpc$, \lymas\ overpredicts
$\xi(s,\mu)$ at large $\mu$ and $s > 2\hmpc$, though it outperforms
the deterministic scheme for $\mu < 0.5$.  It is perhaps somewhat surprising
that the discrepancy persists at large $s$, a point that would be useful
to revisit with hydro simulations of a still larger volume.
The trends in Figure~\ref{fig:correlation_func} agree qualitatively
with those found by \cite{slosar11} in BOSS (their figure 17), but
quantitative assessment will require reproducing the continuum removal
method of Slosar et al., and it probably requires a larger box 
to accurately represent large scale flows.

\section{Creating Pseudo-Spectra}
\label{sec:pseudospectra}

As presented in \S\ref{sec:cpdf}, \lymas\ is a tool for predicting
flux statistics from a dark matter distribution, not for creating
realistic artificial \lyaf\ spectra.
However, it is often useful to create such artificial spectra so that
one can assess the impact of noise, continuum fitting errors, and
other aspects of the observational analysis, and incorporate any
biases from these effects into the model predictions
before comparing to observations.
In this Section we develop a method for creating ``pseudo-spectra''
with realistic line of sight properties that also have
accurate large-scale, 3-d correlations like those predicted 
by \lymas.

\begin{figure*}
\begin{center}
\rotatebox{0}{\includegraphics[width=12.5cm]{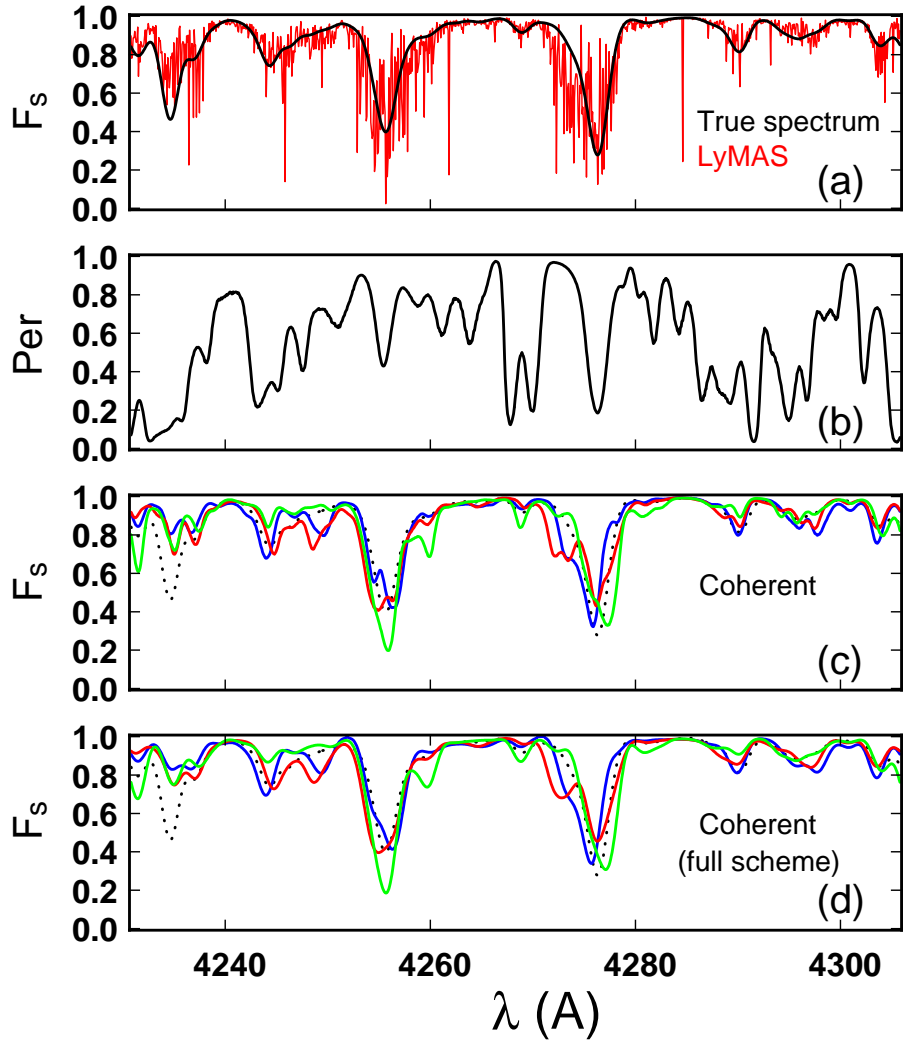}}
\caption{(a) True spectrum along a line of sight through the hydro
simulation (black) compared to a pseudo-spectrum (red) created by
drawing the flux at each pixel independently from the 
conditional distribution $P(F_s|\delta_s)$.
(b) The ``percentile spectrum'' $\per(x)$ for this line of
sight, showing where in the conditional distribution $P(F_s|\delta_s)$
the value of the flux in the true spectrum lies.
(c) Three realizations (with different random number seeds)
of a pseudo-spectrum generated from the dark matter skewer along
this line of sight, using the scheme described in the text but
without the transformation of the 1-d power spectrum and 1-point PDF.
The dotted curve shows the true spectrum.
(d) The same pseudo-spectra {\it after} applying the power spectrum
and PDF transformations described in the text.
 }
\label{fig:spectrum1}
\end{center}
 \end{figure*}

Figure~\ref{fig:spectrum1} illustrates the challenge.
In the upper panel, the black curve shows the true \lyaf\ 
spectrum along a line of sight through the hydro simulation.
The red curve shows the result of taking a skewer through
the smoothed DM density field along the same line of sight,
then drawing flux values at each pixel from our calibrated
conditional distribution $P(F_s|\delta_s)$.
This curve is much ``noisier'' than the true spectrum
because the \lymas\ assumption that separate pixels are
represented by {\it independent} draws from this conditional
distribution obviously breaks down when we are examining
neighboring pixels along the same line of sight.
The second panel of Figure~\ref{fig:spectrum1} demonstrates
this point explicitly by plotting the ``percentile spectrum'':
for each pixel, we plot the fractional position of the flux
value in the true hydro spectrum in the cumulative distribution
of $P(F_s|\delta_s)$.  Specifically, for a pixel with smoothed
flux $F_s$ and smoothed redshift-space density $\delta_s$,
we plot
\begin{equation}
\per(F_s,\delta_{s}) = \int_{0}^{F_{s}} P(F^\prime_s|\delta_s) 
  dF^\prime_s.
\label{eqn:defper}
\end{equation}
(To truly make this a ``percentile'' spectrum we should multiply
by 100, but we instead retain the more natural normalization so 
that $\per$ runs from zero to one.)
We see from Figure~\ref{fig:spectrum1} that when one pixel has a high 
value of $\per(F_s|\delta_s)$, and thus a high flux relative
the conditional mean expected given the local DM density,
then its neighboring pixels also have high values of
$\per(F_s|\delta_s)$.  The coherence of the ``percentile field''
dies away gradually with separation, but independent draws
from $P(F_s|\delta_s)$ lose this coherence entirely.

\begin{figure}
\begin{center}
\rotatebox{0}{\includegraphics[width=\columnwidth]{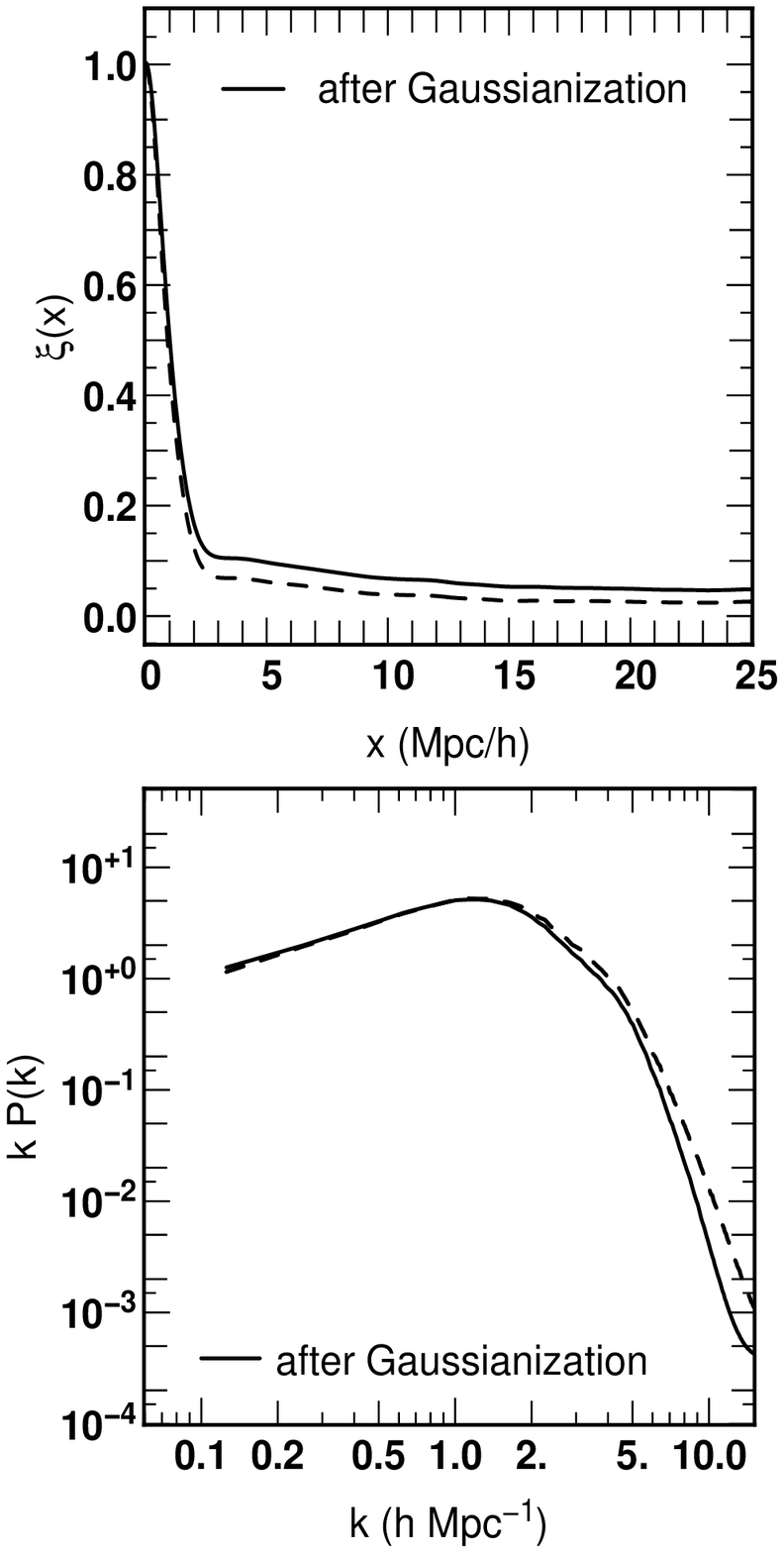}}
\caption{(a) Auto-correlation function of the percentile spectra
computed from the hydro simulation before (dotted) and after 
(solid) Gaussianization.
(b) dimensionless 1-d power spectra of the same cases.  The first mode
plotted is the fundamental of the $50\hmpc$ box.
Both panels show redshift-space results only.
 }
\label{fig:pkper}
\end{center}
 \end{figure}

Figure~\ref{fig:spectrum1} also suggests a strategy for creating
more realistic pseudo-spectra: for each skewer $\delta_s(x)$ through the
smoothed DM density field (where $x$ represents redshift-space position
along the line of sight), first generate a percentile spectrum
$\per(x)$ with the correct statistical properties, then draw the
flux in each pixel from the conditional distribution $P(F_s|\delta_s(x))$
at the location in this distribution implied by $\per(x)$.
The dotted curves in Figures~\ref{fig:pkper}a and~\ref{fig:pkper}b
show, respectively, the line of sight autocorrelation function of 
$\per(x)$ and its Fourier transform, the 1-d power spectrum,
both computed from the full set of 100,000 
spectra from the hydro simulation.
Here we have subtracted the mean $\langle \per(x)\rangle = 0.5$
and divided by the standard deviation 
$\sigma(\per(x)) = \left(\frac{1}{2}+\frac{1}{4}-\frac{1}{3}
  \right)^{-1/2} = 0.6455$
to create a normalized field.
The normalized autocorrelation function drops rapidly from 1.0 at 
zero separation to 0.1 at $x=2\hmpc$, then drops slowly towards
large separations.  The power spectrum is approximately flat
from $k=0.126\,h\,{\rm Mpc}^{-1}$ (the fundamental of the box)
up to $k\approx 1.3\,h\,{\rm Mpc}^{-1}$, then drops rapidly
at higher $k$.

To create percentile spectra, we first Gaussianize $\per(x)$,
applying a monotonic mapping that preserves the rank order of
pixel values but imposes a Gaussian 1-point PDF --- i.e., we
replace $\per(x)$ with
\begin{equation}
G_{\per}(x) = y; \qquad (2\pi)^{-1/2} \int_{-\infty}^y e^{-z^2/2}dz = \per(x).
\label{eqn:Gaussianize}
\end{equation}
Gaussianization was introduced by \cite{weinberg92} as a method
to map the present-day galaxy density field back to cosmological
initial conditions, and it was introduced for \lyaf\ analysis
by \cite{croft98}.  In our case, it ensures that the field
$G_{\per}(x)$ has a Gaussian 1-point PDF, though it does not
ensure that $G_{\per}(x)$ is actually a Gaussian field, which
requires all multi-point PDFs to be multi-variate Gaussians.
Note that we apply the Gaussian mapping collectively to the 
full ensemble of percentile spectra extracted from the hydro
simulation, not one line of sight at a time.
By construction, $G_{\per}(x)$ has zero mean and unit standard deviation.
Solid curves in Figures~\ref{fig:pkper}a and~\ref{fig:pkper}b
show the autocorrelation function and 1-d power spectrum of
this Gaussianized percentile field.  
They differ only moderately from those of $\per(x)$ itself.

For any dark matter skewer,
we can now generate a Gaussian field $g_{\per}(x)$ with same 1-d power spectrum 
as $G_{\per}(x)$ in the usual way, by drawing the real and 
imaginary parts of each Fourier mode $\tilde{g}_k$ from
independent Gaussian distributions with variance $P(k)/2$,
enforcing the Hermitian condition $\tilde{g}_k = \tilde{g}^*_{-k}$,
and Fourier transforming to get a real-valued field $g_{\per}(x)$
(We use lower case $g$ to distinguish these realizations of Gaussian
fields from the original Gaussianized field $G_{\per}$ derived from
the hydro simulation).  After generating $g_{\per}(x)$ fields for
an ensemble of skewers, we ``de-Gaussianize'' them to create
a realization $\lper(x)$ of a percentile field along each line
of sight --- i.e., we monotonically map the values of 
$g_{\per}(x)$ to match the uniform PDF that by definition
describes a percentile field.
(We again use lower case to distinguish our realization of
$\lper(x)$ from the $\per(x)$ field measured from the 
hydro simulation.)
Finally, we create an actual spectrum along each line of sight
by drawing the smoothed flux from $P(F_s|\delta_s)$ at the 
location in the cumulative PDF implied by the realization $\lper(x)$,
assigning each pixel the value of $F_s(x)$ satisfying
\begin{equation}
\int_{0}^{F_s(x)} P(F'_s|\delta_s)dF'_s = 
  (2\pi)^{-1/2} \int_{-\infty}^{g_\per(x)} e^{-z^2/2} dz.
\label{eqn:fluxassignment}
\end{equation}

This procedure is reasonably successful.  The third panel
of Figure~\ref{fig:spectrum1} shows three realizations of spectra
from this procedure along the same line of sight, which can
be compared to the true spectrum in the top panel.
The three realizations differ only in the random number
seed used to generate $\lper(x)$.
In contrast to the red curve in the top panel, these
pseudo-spectra all display a coherence similar to that
of the true spectrum.
In an ensemble of pseudo-spectra, the values of $\per(x)$ will still
be uniformly distributed, and we will therefore sample fairly 
from $P(F_s|\delta_s)$ despite introducing correlations to our draws
for pixels along the same line of sight.
By construction, therefore, an ensemble of pseudo-spectra created
in this way has the same 1-point PDF as the original hydro
spectra and makes the same predictions for the two-point
conditional flux PDFs and flux correlation functions as
the \lymas\ scheme described in \S\ref{sec:cpdf}.
(We have confirmed numerically that we do indeed get identical
results from the two approaches, up to statistical fluctuations
from the finite number of spectra.)
We emphasize that this procedure is {\it not} the same as simply 
generating Gaussian fields and monotonically mapping them to the 
flux PDF of the original hydro spectra, as that approach would
not use the information in the dark matter skewers along each
line of sight and would therefore not build in large scale correlations.

\begin{figure}
\begin{center}
\rotatebox{0}{\includegraphics[width=\columnwidth]{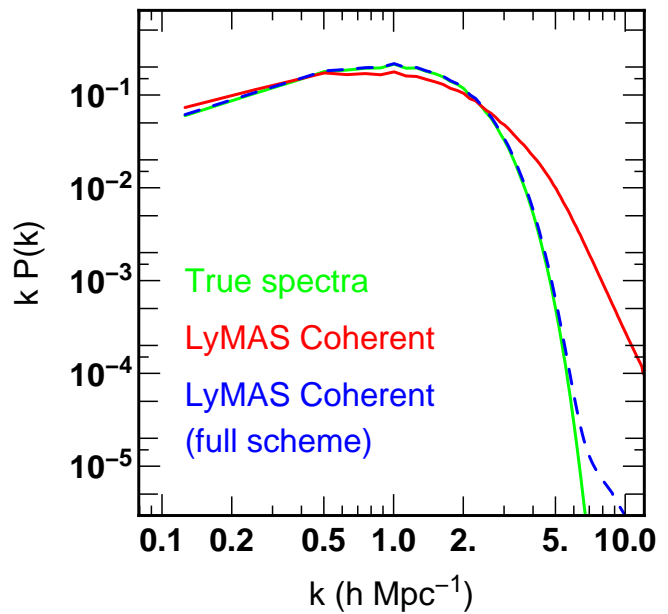}}
\caption{Dimensionless 1-d flux power spectrum of the true spectra from the
hydro simulation (green), the coherent pseudo-spectra 
before power spectrum transformation (red), and the pseudo-spectra
from our full scheme, after applying the power spectrum and
PDF transformations described in the text (blue dashed).
 }
\label{fig:1dpk}
\end{center}
 \end{figure}

Unfortunately, the 1-d power spectrum of pseudo-spectra generated
in this way is not a perfect match to the 1-d power spectrum of
the original hydro spectra.  Figure~\ref{fig:1dpk} shows this
comparison, demonstrating that the pseudo-spectra have more
power on small scales and slightly less power on large scales; 
they are not quite coherent enough.
To solve this problem we borrow a scheme from \cite{weinbergcole92},
who were attempting to create non-Gaussian fields with specified
power spectra via local transformations of Gaussian fields.
In our case, we have a well-defined target power spectrum --- that
of the ensemble of true hydro spectra --- and we have a set of
pseudo-spectra whose 1-d power spectrum $P_{F,PS}(k)$ deviates
from the target $P_F(k)$, primarily at high $k$.  
We repair the discrepancy by computing a 1-d Fourier transform
of each pseudo-spectrum, multiplying each Fourier mode by the
ratio $[P_F(k)/P_{F,PS}(k)]^{1/2}$, and inverse transforming.
The ensemble of modified pseudo-spectra now has the same average
1-d power spectrum as the original hydro spectra, by 
construction.\footnote{We use the values of $P_F(k)$
and $P_{F,PS}(k)$ computed from the ensembles of hydro spectra and
pseudo-spectra, so that we do not artificially force the 1-d 
power spectrum of each individual skewer through the simulation
box to match the same global power spectrum.}
As noted by \cite{weinbergcole92}, this power spectrum modification
alters the 1-point PDF, but the change is small.
For our purposes, we would like to have spectra that have both
the correct 1-d power spectrum and the correct 1-point PDF, and
we therefore apply one more monotonic transformation to the
fluxes of the pseudo-spectra to match the 1-point PDF of the
hydro spectra.  This transformation slightly alters the 1-d
power spectrum, and in principle one could iterate back and
forth between these two steps.  In practice, we find that a
single iteration already produces excellent agreement with 
the target 1-d power spectrum up to quite high $k$, as shown 
by the dotted curve in Figure~\ref{fig:1dpk}.  The 1-point
PDF is perfect by construction.  We therefore stop after this
single iteration.
 We emphasize that our procedure imposes a global match
to the hydro simulation's flux PDF and 1-d $P(k)$ but does not
impose this match spectrum-by-spectrum; each individual spectrum
has the expected random fluctuations about the global mean statistics.

In Figures~\ref{fig:prob_cond_r}-\ref{fig:correlation_func} we have
shown the quantitative predictions of this ``coherent''
\lymas\ scheme with blue dashed curves, and they are 
almost indistinguishable from those of the simpler procedure
described in \S\ref{sec:cpdf}.
Therefore, our more elaborate procedure provides a method
to create ensembles of pseudo-spectra that have the same
1-d power spectrum and 1-point PDF as those from the hydro
simulation and that accurately reproduce the predicted
cross-sightline correlations for transverse separations
larger than $\sim 1\hmpc$.  
Note that we have also verified that the coherent scheme
with $1.0\hmpc$ smoothing generates results similar to
the simple scheme with $1.0\hmpc$.  

\begin{figure}
\begin{center}
\rotatebox{0}{\includegraphics[width=\columnwidth]{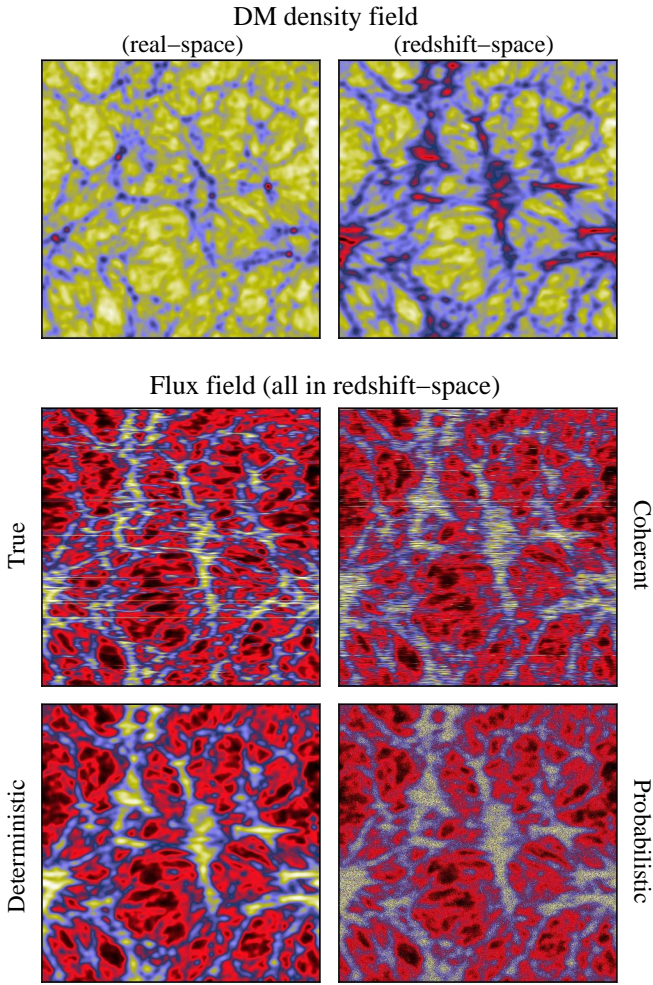}}
\caption{Slices through the smoothed DM density field in real space
(top left) and redshift space (top right) and through the ``flux field''
of the true hydro simulation spectra (middle left) and the flux
fields created from the $\sigma=0.3\hmpc$, redshift-space DM density field using
the deterministic mapping (lower left), the probabilistic mapping
of \S\ref{sec:cpdf} (lower right), and the full coherent method
described in this section (middle right).
The line of sight direction is horizontal in this plot.
Yellow colors correspond to low values, while red and dark are high values;
the top panels are based on logarithm of density.
}
\label{fig:slide}
\end{center}
 \end{figure}

Figure~\ref{fig:slide} illustrates the relation between the 
DM density field and the ``flux field'' $F_s({\bf x})$ that
could in principle be measured on an extremely high density
grid of \lyaf\ sightlines.  Upper panels show slices through
the density field in real space and redshift space,
smoothed with $\sigma = 0.3\hmpc$.  
Lower panels show the flux field in the same slice computed
using the true gas distribution of the hydro simulation,
using the deterministic mapping or the simple \lymas\ 
scheme (here labeled ``probabilistic'') described in 
\S\ref{sec:cpdf}, or using the coherent scheme described in this
section.  It is visually apparent that the true flux field
resembles the redshift-space density field more than the real-space
density field, as one would expect.
In this large scale view, the differences among the three
flux assignment schemes are subtle, though one can see that
the probabilistic scheme loses some of the fine structure
in the true flux field because of its incoherent
sampling of $P(F_s|\delta_s)$.

We close this section with a technical recap of the full \lymas\ 
recipe for creating an ensemble of pseudo-spectra from a dark matter
simulation.  The inputs needed from the hydro simulation are:\hfill\break
1. The conditional flux distributions $P(F_s|\delta_s)$ for whatever 3-d
smoothing length will be applied to the DM simulation.  For BOSS
resolution spectra, we find that $\sigma=0.3\hmpc$ is an optimal
choice, but $\sigma=1.0\hmpc$ still yields good results for
transverse separations above $\approx 2\hmpc$.
\hfill\break
2. The 1-d power spectrum and unconditional PDF of the hydro spectra.
\hfill\break
3. The 1-d power spectrum of the Gaussianized percentile spectra $G_{\per}(x)$.
\hfill\break
In the website
dedicated to LyMAS,
 we provide
tabulations of these quantities derived from the RAMSES simulation
so that others can use these calibrations to create pseudo-spectra
from their own DM simulations.

With these inputs, the steps for creating pseudo-spectra are:\hfill\break
1. Extract skewers through the smoothed DM density field.
Our tests in this paper are based on the fairly sophisticated smoothing
scheme described in \S\ref{sec:simulation}, and we have not tested
whether simpler methods give adequate results.
\hfill\break
2. For each DM skewer, create a realization $g_{\per}(x)$ of a
1-d Gaussian field with the hydro-calibrated power spectrum of
$G_{\per}(x)$.  The $P(k)$ computed from our $50\hmpc$ hydro 
simulation necessarily cuts off at $k=2\pi/(50\hmpc)$, but the
coherence scale of $G_{\per}(x)$ is clearly much smaller than
this, about $2\hmpc$ in the autocorrelation function 
(Fig.~\ref{fig:pkper}), so we recommend smoothly truncating
the power to zero for the lower $k$ values that arise in larger
simulations.  
\hfill\break
3. Along each skewer, create an initial flux field $f_0(x)$ by
drawing the flux at each pixel from the location in $P(F_s|\delta_s)$
implied by the value of $g_{\per}(x)$ (eq.~\ref{eqn:fluxassignment}).
\hfill\break
4. Measure the 1-d flux power spectrum $P_{f_0}(k)$ of the ensemble
of pseudo-spectra created in this way.  Then Fourier transform
each pseudo-spectrum, multiply each of its Fourier components
by the ratio $[P_F(k)/P_{f_0}(k)]^{1/2}$, and inverse transform
to get modified pseudo-spectra $f_1(x)$ that have the same 1-d
flux power spectrum $P_F(k)$ as that from the true hydro spectra.
\hfill\break
5. Monotonically map the flux values of $f_1(x)$ so that
their unconditional PDF matches that of the hydro spectra, 
obtaining the final realization of the smoothed flux spectra $F_s(x)$.

While one could iterate steps 4 and 5, in our test the final 
monotonic mapping barely alters the 1-d flux power spectrum, so an
iteration is unnecessary.  For some purposes, one may prefer
pseudo-spectra calibrated to match observations rather than 
representing the prediction of a theoretical model.
In this case, one could use empirical estimates of the flux power
spectrum and unconditional PDF in place of the hydro simulation
results in steps 4 and 5, while still using the DM simulation to
build in realistic large scale flux correlations.  One could potentially
use an approximate method in place of an N-body simulation
to generate the redshift-space DM density field, such as a log-normal
transformation of a linear theory, Gaussian realization.
One caveat is that this density field will not have the same density
PDF as that of the hydro simulation used for calibration, so the
pseudo-spectra created in step 3 will not have the correct
flux PDF.  The results after steps 4 and 5 may still be good,
but we have not tested such a case.

\section{Application to large cosmological dark matter simulations}
\label{sec:appli}

\begin{figure}
\begin{center}
\rotatebox{0}{\includegraphics[width=\columnwidth]{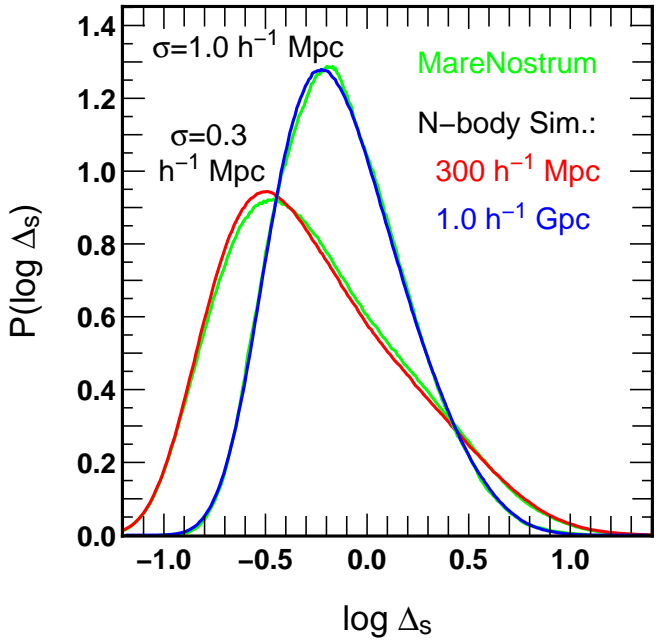}}
\caption{Probability distribution functions of $\log\Delta_s$, the
logarithm of the smoothed dark matter overdensity,
from our $1024^3$ Gadget-2 N-body simulations of a $300\hmpc$ box
(red) and a $1.0\hgpc$ box (blue), with 3-d smoothing lengths
$\sigma=0.3\hmpc$ and $\sigma=1.0\hmpc$, respectively. 
Green curves show the dark matter overdensity PDFs from the MareNostrum
simulation with the same smoothing lengths.
All results are at $z=2.51$.
}
\label{fig:dm_pdf}
\end{center}
 \end{figure}

In this section we apply our full \lymas\ 
recipe for creating an ensemble of pseudo-spectra 
from a large volume dark matter simulation. 
However, 
our calibrations of $P(F_s|\delta_s)$ are based on the dark matter
distribution of a high-resolution hydro simulation.
Two potential complications in applying \lymas\ to a larger volume
dark matter simulation are that the dark matter in the hydro simulation
is affected by gravity of the dissipative baryons and that the PDF of the
smoothed dark matter field may be affected by lower resolution.
We expect the first effect to be small at the smoothing scales
and moderate overdensities relevant to the \lya\ forest.  To address the
resolution question we ran
several cosmological N-body simulations with Gadget2 (\citealt{springel05})
with different box lengths and similar cosmological parameters to those used
in the MareNostrum simulation.
We found that the PDF of the smoothed dark matter density contrast
$P(\delta_s)$ agrees well with that from the MareNostrum simulation if
the mean inter-particle distance $n_p^{-1/3}$ 
is similar to the considered smoothing length,
as illustrated in Figure~\ref{fig:dm_pdf}.



In the following, we analyze two cosmological dark matter simulations performed with Gadget2.
 In order to produce and analyze a reasonable
amount of data, each simulation is run with $1024^3$ dark matter particles. 
Thus,  to apply our scheme for  $\sigma=1\hmpc$ and
$\sigma=0.3\hmpc$, appropriate box lengths are 
respectively  $1\hGpc$ and $300\hmpc$.
Initial conditions have been generated using the 
MPgrafic code (\citealt{prunet08}). The  $1\hGpc$ and $300\hmpc$ simulations 
start at $z=18.8$ and $z=30.4$, respectively, and end at
$z=2.51$ where we create our ensembles of pseudo-spectra.
In each simulation, the Plummer-equivalent force softening adopted
is 5\% of the mean inter-particle distance
($48.8\hkpc$ and $14.6\hkpc$ respectively),
kept constant in comoving units.
We generate redshift-space dark matter density fields on a $4096^3$ grid.
While the PDFs of $\Delta_s = 1+\delta_s$ are already close
to those obtained from the hydro simulation (Figure~\ref{fig:dm_pdf}),
we monotonically remap $\Delta_s$ to enforce perfect agreement
before generating our pseudo-spectra.

\begin{figure}
\begin{center}
\rotatebox{0}{\includegraphics[width=\columnwidth]{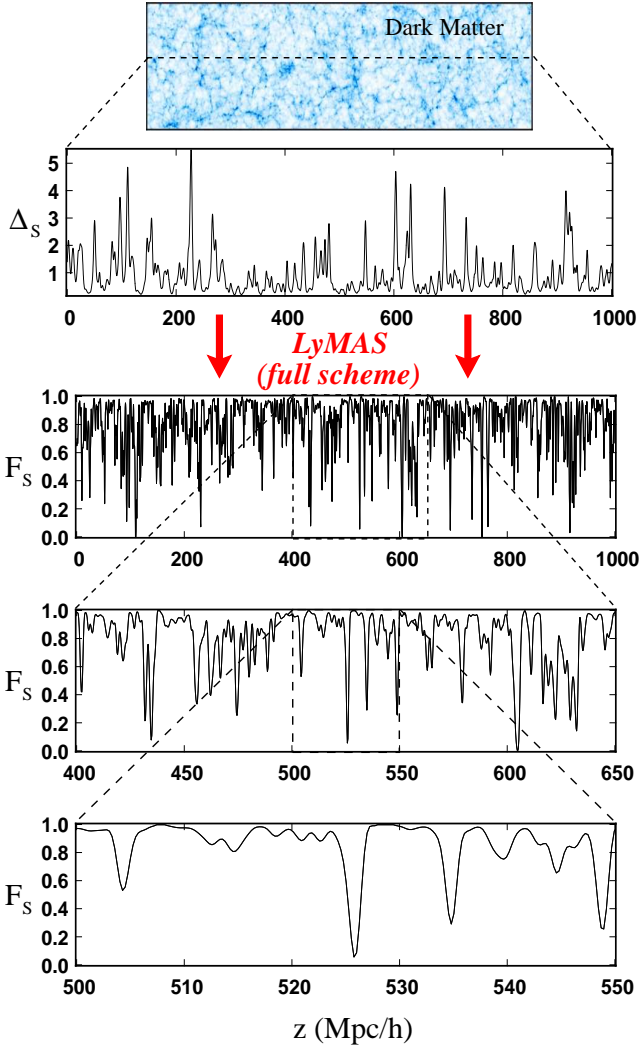}}
\caption{Application of the full \lymas\ scheme
to a large, pure dark matter simulation ($1\hGpc$ - $1024^3$ particles).
The second panel plots the dark matter overdensity along the skewer
indicated in the top panel, and the three lower panels show
successive zooms on the \lymas\ pseudo-spectrum, with redshift
path lengths (from top to bottom) of $\Delta z = 1.23$, 0.185,
and 0.062.  Equivalently, for the bottom panel, $\Delta\lambda = 75$\AA\ 
and $\Delta v = 5271\kms$.  
}
\label{fig:bigbox}
\end{center}
 \end{figure}

\begin{figure}
\begin{center}
\rotatebox{0}{\includegraphics[width=\columnwidth]{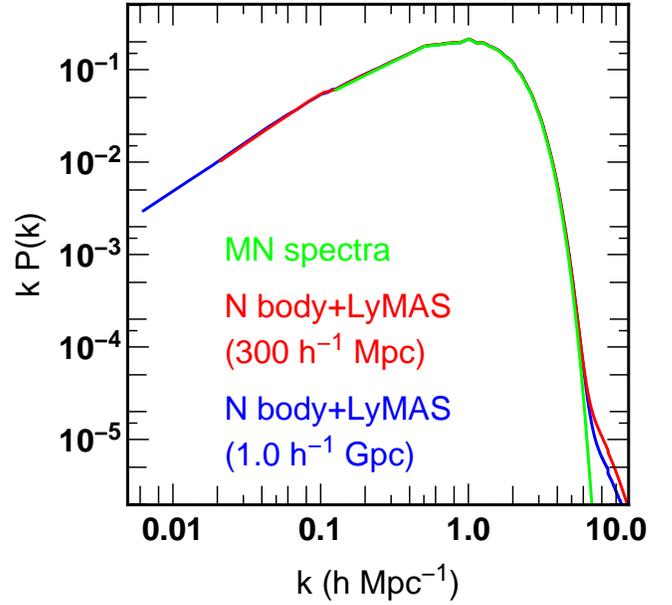}}
\caption{The dimensionless, redshift-space, 
1-d flux power spectrum of the true spectra from the
hydro simulation (green) and 
coherent pseudo-spectra (full scheme with one iteration)
from the $300\hmpc$ (red) and  $1.0\hGpc$ (blue) box side N-body simulations. 
}
\label{fig:Pk_bigbox}
\end{center}
 \end{figure}

\begin{figure}
\begin{center}
\rotatebox{0}{\includegraphics[width=\columnwidth]{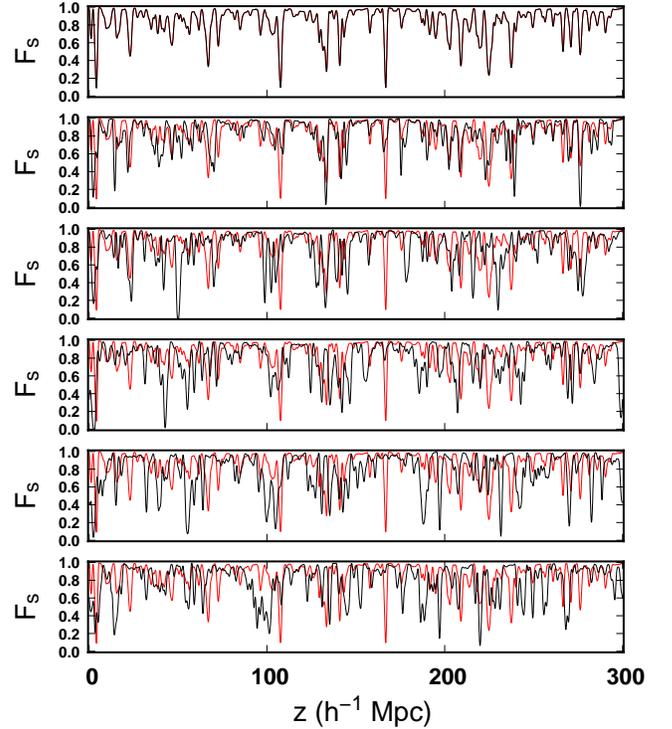}}
\caption{Sample of six pseudo-spectra extracted from the $300\hmpc$ box
simulation along parallel line of sight at  $1\hmpc$ intervals in transverse
separation.  The top spectrum is repeated as a red curve in
the subsequent panels.
}
\label{fig:stack1}
\end{center}
 \end{figure}

Figure~\ref{fig:bigbox} illustrates
a pseudo-spectrum through the $1\hgpc$ simulation.
The full line of sight has a redshift path length
$\Delta z = 1.23$, comfortably larger than the 
$\Delta z = 0.55$ between the \lya\ and Ly$\beta$ emission
lines of a $z=2.5$ quasar.
The lowest, zoomed panel is the same length as a single 
line of sight through the MareNostrum box, corresponding
to an observed-frame wavelength range of 75\AA.
Figure~\ref{fig:Pk_bigbox} compares the ensemble-average 1-d
power spectrum of the MareNostrum hydro spectra and the \lymas\ 
pseudo-spectra from the $300\hmpc$ and $1.0\hgpc$ box.
By construction, the power spectra agree essentially perfectly
over the common range, with the larger boxes continuing
out to lower $k$.  Because we use only a single iteration,
there are discrepancies at $k \geq 0.6\hmpc$, but the power on
these scales is too weak to noticeably affect the flux spectra.
We have also confirmed that the unconditional flux PDFs of the three
sets of spectra match perfectly.  Figure~\ref{fig:stack1} 
shows six spectra extracted from the $300\hmpc$ box at $1\hmpc$
intervals of transverse separation.  Across a $1\hmpc$ separation,
the coherence of structure along neighboring lines of sight
is visually evident.  By $5\hmpc$, there is little coherence
still obvious to the eye, though flux statistics do reveal
correlations at these scales and larger.

\begin{figure*}
\begin{center}
\rotatebox{0}{\includegraphics[width=16cm]{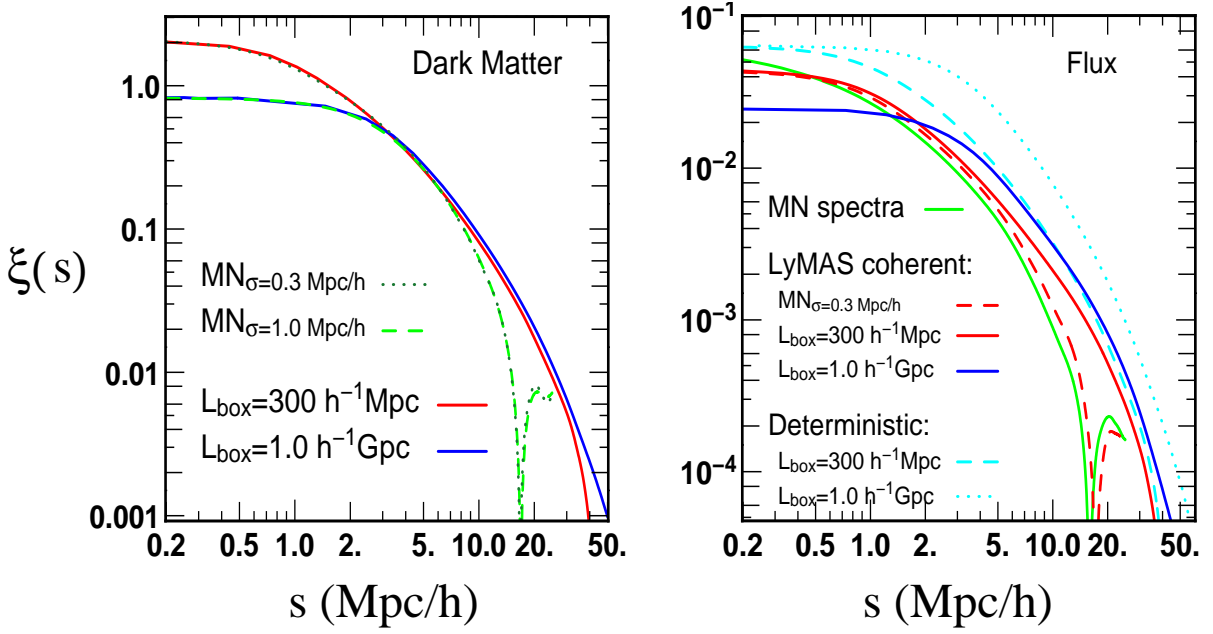}}
\caption{(Left) Angle-averaged redshift-space correlation function of 
the smoothed dark matter density fields,
$\xi=
\langle \Delta_{s}(r,z) \Delta_{s}(r+dr,z+dz)\rangle -1$, 
as a function
of the separation $s=(dr^2+dz^2)^{1/2}$,
derived from the $50\hmpc$ MareNostrum simulation (dashed and dotted green lines)
and the $300\hmpc$ (red) and $1.0\hmpc$ (blue) N-body simulations.
The MareNostrum correlation function is negative beyond
$s=16\hmpc$, a consequence of missing power on scales larger
than the box.  
The smoothing length is $\sigma=0.3\hmpc$ for MareNostrum (dotted green line) and
the $300\hmpc$ box, and $\sigma=1.0\hmpc$ for MareNostrum (dashed green line) and the $1.0\hgpc$ box.
(Right) Angle-averaged redshift-space correlation function 
of the flux, 
$\xi_F= \langle F(r,z) F(r+dr,z+dz)\rangle/ \langle F \rangle^2 -1$,
for spectra at BOSS resolution.
The green curve shows  results from the hydro spectra in the 
MareNostrum simulation.  The dashed red curve shows results from \lymas\
applied to the MareNostrum dark matter distribution, with 
$\sigma = 0.3\hmpc$.  Red and blue curves show \lymas\ predictions
from the $300\hmpc$ and $1.0\hgpc$ simulations with 
$\sigma = 0.3\hmpc$ and $1.0\hmpc$, respectively.
Dashed and dotted cyan curves show predictions of the deterministic
scheme applied to these simulations, with the same smoothing lengths.
We consider the red curve to be the most realistic prediction,
since it has the high accuracy of the $\sigma=0.3\hmpc$ \lymas\ 
method but does not suffer from the strong box size effect of the
MareNostrum simulation.
}
\label{fig:big_corr_dm}
\end{center}
\end{figure*}

The left panel of Figure~\ref{fig:big_corr_dm} shows the correlation
functions (3-d, redshift-space) of the smoothed dark matter density
fields from the MareNostrum simulation and from the $300\hmpc$
and $1.0\hgpc$ N-body simulations.  The correlation function of
the $50\hmpc$ MareNostrum box is noticeably suppressed at $s=10\hmpc$
and driven to zero at $s=16\hmpc$.  The correlation functions of
the N-body simulations remain positive to $s\geq 30\hmpc$.
In this plot, the results from the $1.0\hgpc$ simulation should
be the most realistic because they are least affected by cutoff
of power at the box scale.  

The right panel of 
Figure~\ref{fig:big_corr_dm} 
shows the redshift-space flux correlation function $\xi_F(s)$,
averaged over angles and 
excluding pixel pairs along the same line of sight.  
The correlation function of the MareNostrum spectra goes negative at $s\approx 16\hmpc$,
reflecting the behavior of the dark matter correlation function
in the left panel.
The dashed red line in this panel show the flux correlation function computed
from \lymas\ pseudo-spectra in the MareNostrum box, with 
$\sigma=0.3\hmpc$.  This accurately tracks $\xi(s)$ from the true hydro
spectra, as expected based on the previous results in 
Figure~\ref{fig:correlation_func}.  The red curve shows $\xi(s)$
computed by applying \lymas\ to the $300\hmpc$ box, again with
$\sigma=0.3\hmpc$; $\xi(s)$ now remains positive to $s \geq 30\hmpc$.
The blue curve shows $\xi(s)$ from the $1.0\hgpc$ box with $\sigma=1.0\hmpc$.
While the larger simulation has better representation of large scale
power, the predictions are not necessarily more realistic,
because $1.0\hmpc$ smoothing yields less accurate results for $\xi(s)$
than $0.3\hmpc$ smoothing (Fig.~\ref{fig:correlation_func}), and
the smoothing effect is likely more important than the box size
effect on these scales.  By contrast, the prediction of the $300\hmpc$
simulation is almost certainly more realistic than that of the
MareNostrum simulation, as \lymas\ with $\sigma=0.3\hmpc$ yields
accurate predictions of $\xi(s)$ and the box size effects are
much weaker for $300\hmpc$ than for $50\hmpc$.

Dotted and dashed cyan curves in this panel show
results from applying the deterministic mapping prescription to the
two simulations, with the same smoothing lengths.  In each case,
the deterministic mapping significantly overpredicts the flux
correlation function relative to the more accurate \lymas\ scheme,
as seen previously for specific $\mu$ values in 
Figure~\ref{fig:correlation_func}.
This difference is not surprising, as the stochasticity in the
relation between density and flux should suppress the correlation
function of the latter with respect to the former.
It is important to take these stochastic effects into account
when predicting large scale flux correlations from a cosmological model.

\begin{figure}
\begin{center}
\rotatebox{0}{\includegraphics[width=\columnwidth]{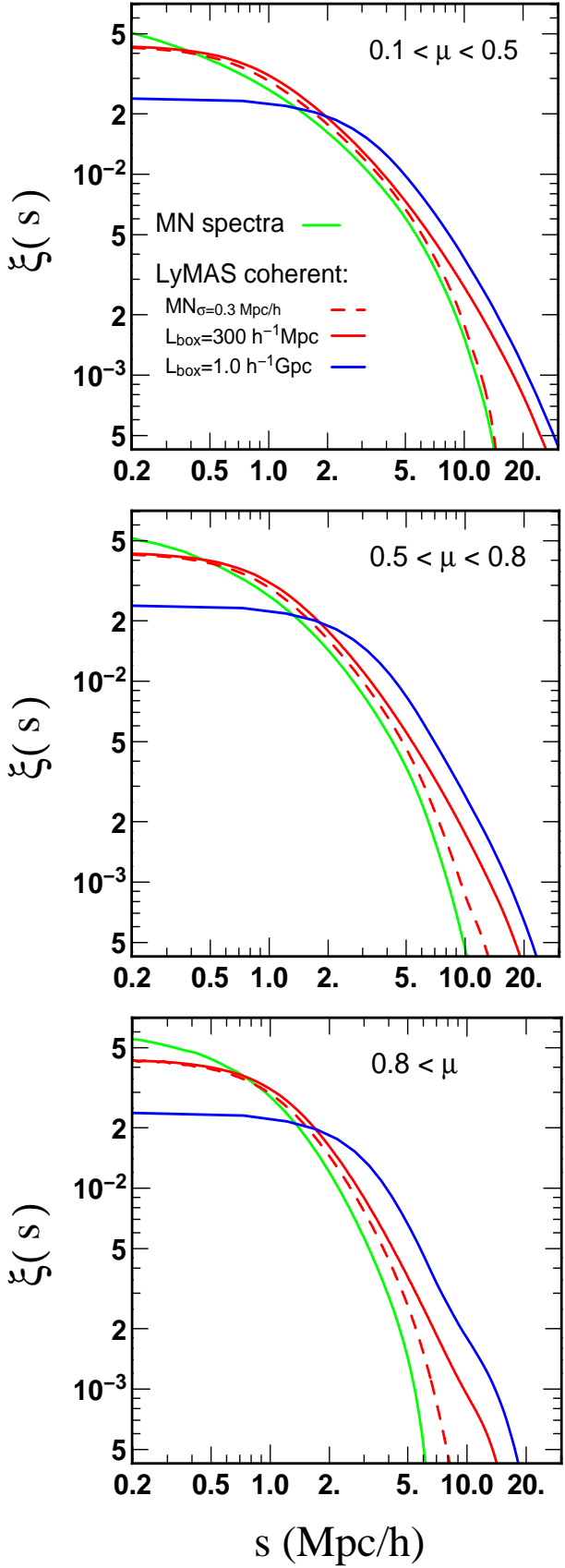}}
\caption{
Flux correlation functions averaged over bins of angle $\mu$,
as labeled.  In each panel, green curves show results from the
hydro spectra of the MareNostrum simulation, dashed red curves show
results from applying \lymas\ to the MareNostrum simulation
with $\sigma = 0.3\hmpc$, and red and blue curves show \lymas\ 
results from the $300\hmpc$ box ($\sigma=0.3\hmpc$) and the
$1.0\hgpc$ box ($\sigma=1.0\hmpc$), respectively.
}
\label{fig:big_corr_F}
\end{center}
 \end{figure}

Figure~\ref{fig:big_corr_F} plots the
flux correlation function averaged in three broad bins of
the angle $\mu$, defined as described at the end of 
Section~\ref{sec:cpdf}.
The amplitude of $\xi_F(s,\mu)$ decreases with increasing $\mu$
at fixed $s$, as seen previously in Figure~\ref{fig:correlation_func}.
Comparison of the red lines and the dashed red ones shows the influence
of box size between the $50\hmpc$ and $300\hmpc$ simulation volumes.
As shown previously in Figure~\ref{fig:correlation_func}, \lymas\ is
less accurate for high values of $\mu$, indicating some shortcomings
at modeling redshift-space distortion for lines of sight with small
transverse separations.  Nonetheless, the $300\hmpc$ box
predictions are likely more realistic than those of the 
MareNostrum hydro spectra because of the larger simulation volume
and the good but not perfect accuracy of \lymas.  The higher
amplitude correlation function of the $1.0\hgpc$ box could be
partly due to its larger simulation volume, but comparison
to Figure~\ref{fig:correlation_func} suggests that the dominant
effect is the lower accuracy of \lymas\ with $1.0\hmpc$ smoothing.

\begin{figure*}
\begin{center}
\rotatebox{0}{\includegraphics[width=16cm]{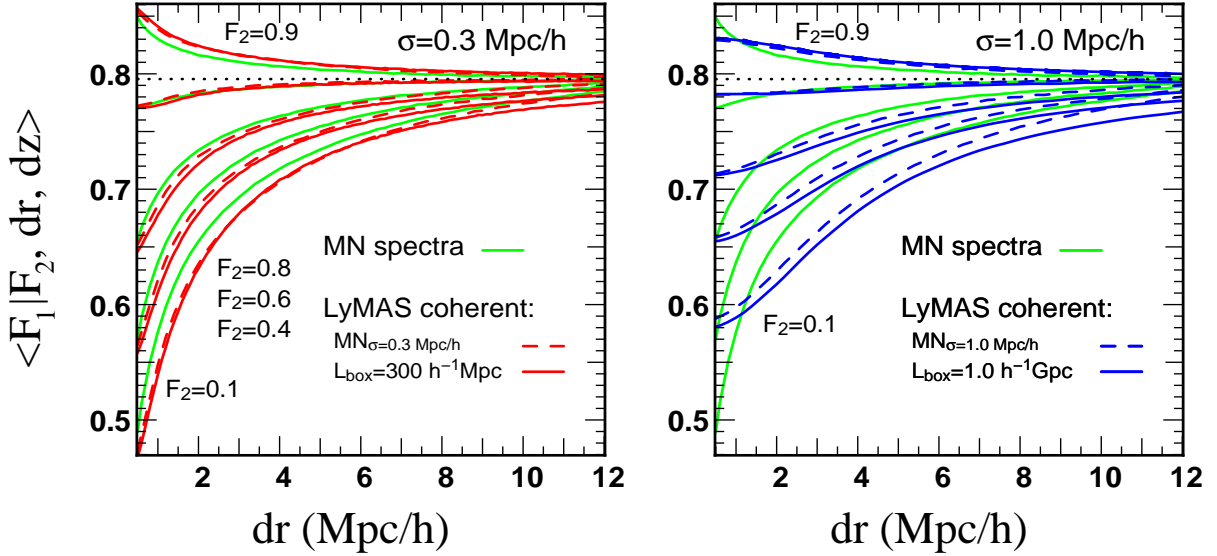}}
\caption{Conditional mean flux $<F_1|F_2,dr,dz>$ 
as a function of transverse separation $dr$ (with $dz=0$),
for the labeled values of $F_2$.
Green curves are derived from the hydro spectra in the 
$50\hmpc$ MareNostrum box. Dashed red curves and dashed blue curves show
results from applying \lymas\ to the MareNostrum simulation
with $\sigma = 0.3\hmpc$ (left panel) and $\sigma = 1.0\hmpc$ (right panel),
respectively.
Curves show \lymas\ results from the $300\hmpc$ box
(left panel, with $\sigma=0.3\hmpc$) and the $1.0\hgpc$ box
(right panel, with $\sigma=1.0\hmpc$).
}
\label{fig:big_prob_cond}
\end{center}
 \end{figure*}

\begin{figure}
\begin{center}
\rotatebox{0}{\includegraphics[width=\columnwidth]{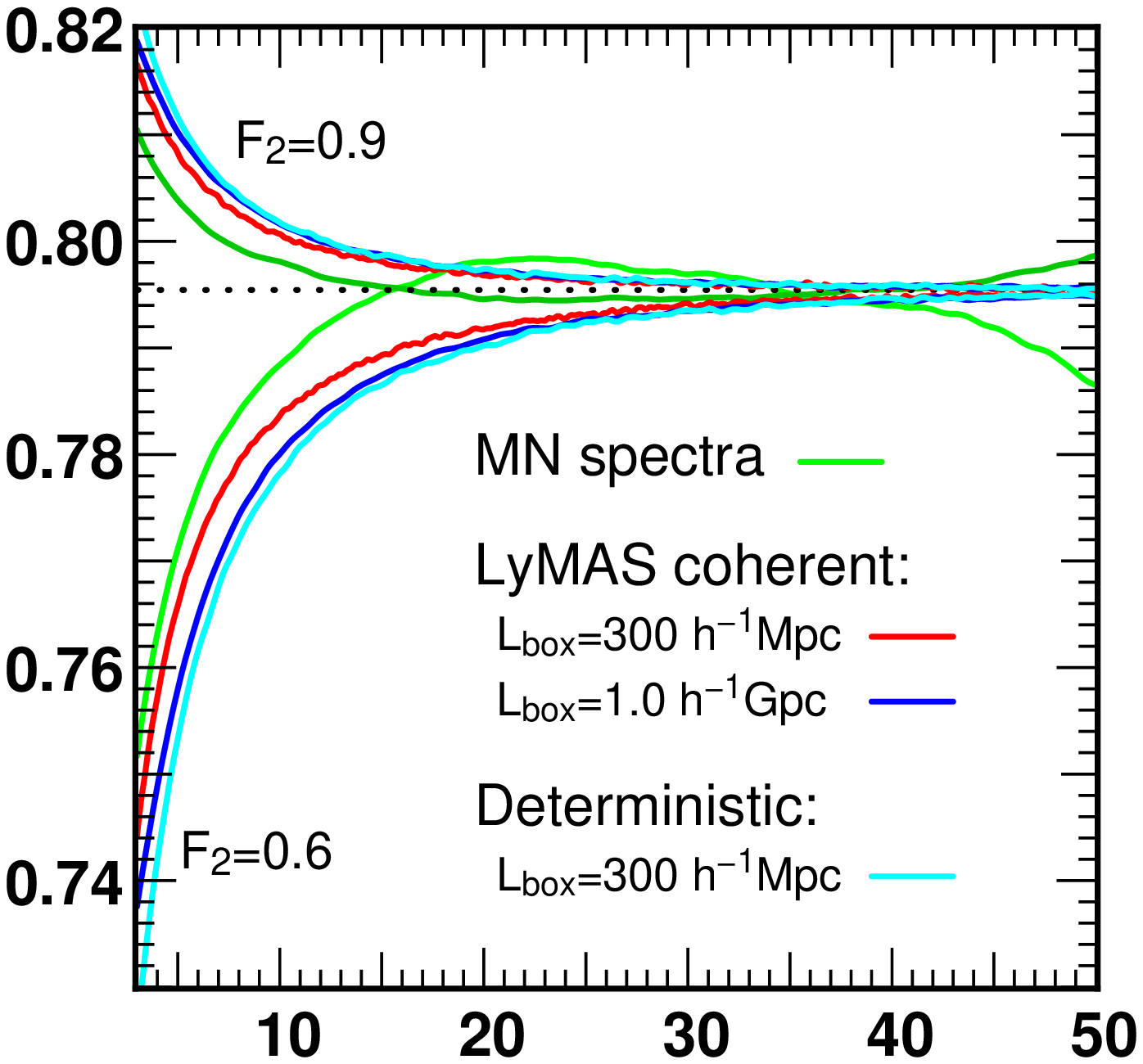}}
\caption{Conditional mean flux at larger separation,
$dr = 3-50\hmpc$, for $F_2 = 0.9$ and 0.6.
We show results from the hydro spectra of the MareNostrum box
and from \lymas\ and deterministic mapping applied to the
$300\hmpc$ box with $\sigma=0.3\hmpc$.
}
\label{fig:big_prob_cond2}
\end{center}
 \end{figure}

Figure~\ref{fig:big_prob_cond} shows the conditional mean
flux $<F_1|F_2,dr,dz>$ 
as a function of transverse separation $dr$ (with $dz=0$),
for several values of the conditioning flux $F_2$.
As shown previously in Figure~\ref{fig:mean_cond_f},
\lymas\ with $\sigma=0.3\hmpc$ reproduces the full hydro
predictions for this statistic almost perfectly for $dr \geq 1\hmpc$
in the MareNostrum box.  The difference between dashed red curves
and red curves in the left panel therefore reflects the larger
box size of the $300\hmpc$ simulation, which noticeably
increases the signature of flux correlations on this statistic
even for $dr$ as small as $2-5\hmpc$.
The right panel shows results from the $1.0\hgpc$ box,
but the greater separation between dashed blue and blue curves
in this panel likely reflects the lower accuracy
of \lymas\ with $\sigma=1.0\hmpc$ (see Fig.~\ref{fig:mean_cond_f}).
Figure~\ref{fig:big_prob_cond2}
shows the conditional mean decrement at separations $dr = 5-50\hmpc$,
for the $50\hmpc$ and $300\hmpc$ boxes only. 
The conditional signal has vanished by $dr=10\hmpc$ in the
MareNostrum spectra, but in the $300\hmpc$ box it continues out
to $dr = 25\hmpc$.  Spectra created by deterministic mapping
in the $300\hmpc$ box also show this larger scale coherence,
but they overpredict the magnitude of the conditional mean effect.

We are making the catalogs of mock spectra from the $300\hmpc$
and $1.0\hgpc$ simulations
publicly available\footnote{http://www2.iap.fr/users/peirani/lymas/lymas.htm}, as they may be broadly useful for investigating
statistics of the 3-d \lya\ forest.
We also make hydro and \lymas\ coherent spectra from the 
MareNostrum box available so that others can look at the accuracy
 of \lymas\ for statistics they are interested in.
These spectra are at $z = 2.51$, and they represent the
cosmological model and IGM thermal history used in the MareNostrum
simulation.  They are smoothed to BOSS spectral resolution, and
they are noiseless.  We defer the creation of mock spectra that
incorporate redshift evolution and noise, and variations of cosmological
parameters or IGM parameters, to future work.
By construction, these mock spectra reproduce the 1-d flux
power spectrum and flux PDF predicted by the MareNostrum
hydro simulation.  For predictions of 3-d flux statistics,
the spectra from the $300\hmpc$ box should yield similar predictions
to those that would be obtained from a RAMSES simulation
$(300/50)^3 = 216$ times larger  in volume.
Predictions from the $1.0\hgpc$ box suffer from the lower 
accuracy of \lymas\ with $\sigma=1.0\hmpc$ smoothing.
Nonetheless, these predictions should be significantly more
accurate than those from a deterministic flux mapping scheme,
and thus from any previous simulation of the \lya\ forest
over such a large volume.

\section{Conclusions and Outlook}
\label{sec:discussion}

\lymas\ is a promising method for making accurate predictions for
flux statistics of the 3-d \lya\ forest on intermediate and large
scales.  The underlying idea is that flux correlations on these
scales are driven by correlations of the dark matter density field,
so that flux statistics can be computed by combining the dark matter
density field with the conditional PDF of flux given density,
$P(F_s|\delta_s)$.  Here we have calibrated these conditional 
PDFs using the $50\hmpc$ MareNostrum AMR hydrodynamic simulation,
focusing on spectra matched to the resolution of BOSS at $z \approx 2.5$.
We find the tightest flux-density correlation and the best overall
performance of \lymas\ for a 3-d dark matter smoothing length
$\sigma \approx 0.3\hmpc$ (with similar performance down to 
smaller smoothing lengths of $\sigma \approx 0.2\hmpc$).
The conditional PDFs are quite broad even for this smoothing
length, but \lymas\ nonetheless gives accurate predictions for the 
two-point flux distribution and correlations down to transverse
separations as small as $2\hmpc$.  It is substantially more
accurate than predictions from a deterministic flux-density mapping,
even when the latter is optimally tuned to reproduce the unconditional
flux PDF of the hydro simulation.  Redshift-space distortions have
a different impact on the matter distribution and the \lya\ forest,
because mass is conserved under the peculiar velocity mapping while
absorption is not.  Nonetheless, \lymas\ predictions for redshift-space
flux correlations from the redshift-space dark matter field prove
just as accurate as predictions for the real-space flux correlations
(i.e., with peculiar velocities set to zero) from the real-space
density field.

For forward modeling tests it is desirable to create artificial
spectra that can be analyzed with the same procedures applied to
observational data.  By forming the ``percentile fields''
described in \S\ref{sec:pseudospectra}, we are able to make
spectra that are coherent from pixel to pixel along a single
line of sight and have the same cross-sightline statistics as
the simple \lymas\ method in which each pixel flux is drawn
independently from $P(F_s|\delta_s)$.
However, the 1-d flux power spectrum of
these coherent spectra has more high-$k$ power than that measured
from the full hydro spectra.  By remapping first the amplitudes
of Fourier modes and then the pixel flux values, 
as described in \S\ref{sec:pseudospectra}, we are able
to create model spectra that exactly reproduce the 1-d flux power 
spectrum {\it and} the unconditional flux PDF of the full hydro
spectra.  The remapping slightly alters the cross-sightline
statistics, but they appear to become, if anything, more accurate
in reproducing the full hydro results.

For practical purposes it would be useful to adopt larger dark matter
smoothing lengths for large scale predictions, so that \lymas\ could
be applied to lower resolution dark matter simulations of very
large volumes.  Unfortunately, \lymas\ with $\sigma=1.0\hmpc$
proves noticeably less accurate than \lymas\ with $\sigma=0.3\hmpc$
even on scales that are much larger than either smoothing length,
tending to overpredict the strength of flux correlations.
It is still significantly more accurate than a deterministic
mapping, even when the latter is applied at $\sigma=0.3\hmpc$,
so \lymas\ remains the best method to 
make mock \lya\ forest spectra from a low resolution
N-body simulation.  However, $\sigma = 0.3\hmpc$ seems required
to achieve the full accuracy of the method, which in turn requires
an N-body simulation with $n_p^{-1/3} \la 0.3\hmpc$.
We have not investigated whether a smaller dark matter
smoothing length would be needed to model high-resolution spectra (from KECK/UVES
for instance).
We have also not investigated whether larger smoothing lengths
would be acceptable, or even preferred, for modeling spectra with
lower resolution than BOSS (X-SHOOTER for instance).
But in all cases, one would still need to generate new calibrations
$P(F_s|\delta_s)$ and new power spectrum of percentile spectra.

\lymas\ shifts the problem of simulating Gpc$^3$
volumes of the \lya\ forest from the realm of computationally
impossible to computationally challenging.  An N-body simulation
of a $1\hgpc$ cube with high enough resolution to produce 
$\sigma=0.3\hmpc$ dark matter skewers requires $\sim 10^{10}$
particles, and even the creation of the dark matter field
from the N-body particle distribution is a cpu- and memory-intensive task.
One also requires a calibrating hydrodynamic simulation for each
cosmological model of interest, of large enough volume to yield 
accurate $P(F_s|\delta_s)$.  As a first step, we have created
catalogs of artificial spectra at BOSS resolution for
the MareNostrum cosmology at $z = 2.5$, from a $300\hmpc$
box with $\sigma = 0.3\hmpc$ and a $1\hgpc$ box with 
$\sigma=1.0\hmpc$.  In future work, we will investigate methods
for incorporating the redshift evolution of the forest 
over the range probed by an individual quasar spectrum and
move towards a full realization (from multiple simulations)
of the BOSS quasar sample.

Alongside these technical challenges, the most interesting
direction for future study with \lymas\ is to explore the sensitivity of medium
and large scale structure in the 3-d \lya\ forest to the
underlying cosmological and IGM parameters.
 Indeed, if the thermal history
of the hydro simulation or the cosmological parameters change, the PDF of the flux
and then the probabilities $P(F_s|\delta_s)$  would also change. At present, we don't know 
if we still need to derive new calibration tables for each set of specific cosmological
parameters (or set of thermal history models) or whether a simple rescaling at some 
steps from the present study would be acceptable.  Nevertheless, 
These investigations can be started by manipulating the
MareNostrum simulation to compute conditional flux PDFs
for different IGM equations of state (by artificially changing
gas temperatures) or matter fluctuation amplitudes (by 
relabeling output redshifts), then applying these conditional PDFs
to different N-body outputs.  These rescaling calculations
should be spot-checked against fully self-consistent
sets of simulations, but we expect they will provide good
guidance to the sensitivity of flux statistics to cosmology
and degeneracies with the IGM.
There are obvious analogies between
the \lymas\ approach to modeling the \lya\ forest and methods
for creating galaxy redshift catalogs by populating the dark 
matter halos of N-body simulations in accord with semi-analytic
models of galaxy formation, with the advantage that predicting
the \lya\ forest from first principles is more straightforward
than modeling galaxy formation.
However, the effects of ionizing background fluctuations
\citep{croft04}, inhomogeneous reionization \citep{mcquinn11}, 
and metal-line contamination
remain important potential systematics that require
further scrutiny.  

The potential payoff from exploiting sub-BAO scales in the
3-d \lya\ forest is large.  One long-recognized opportunity
is to apply the Alcock-Paczynski (\citeyear{alcock79}) test to the
\lya\ forest to constrain the product $H(z)D_A(z)$ at
high redshift \citep{hui99,mcdonald99}; this requires accurate
predictions of the intrinsic anisotropy in \lya\ forest clustering.
A more model-dependent variant of this idea is to use the amplitude
of correlations as a function of transverse separation to constrain
$D_A(z)$ directly.  The slope and curvature of the matter
power spectrum are diagnostics for neutrino masses or 
curvature in the inflationary
fluctuation spectrum, and the \lya\ forest can probe 
the power spectrum shape on scales that
are degraded by non-linear evolution at low redshift.
Most enticing but perhaps most challenging is the prospect
of precisely constraining the amplitude of matter clustering
at $z = 2 - 4$, so that $\Lambda$CDM predictions for the growth
of structure can be tested from the CMB epoch through the
\lya\ forest epoch down to the lower redshifts probed by large
weak lensing and galaxy redshift surveys.  
These tests could reveal the signatures of complex dark
energy models or devations from General Relativity on 
cosmological scales.  With the massively multi-plexed spectroscopic
surveys underway and planned for the future, the 3-d \lya\ forest
is emerging as a major new route to tracing the growth of
structure in the universe.  \lymas\ is an important step
towards matching theoretical predictions to this
observational opportunity.

\acknowledgments
\noindent
DW acknowledges the hospitality of the Institut d'Astrophysique de Paris,
support of grant AST-1009505 from the U.S. National Science Foundation,
and additional support from French state funds managed by the ANR within the
Investissements d'Avenir program under reference ANR-11-IDEX-0004-02.
We also  acknowledge support from the ``
Programme National Cosmologie et Galaxies''. 
This work was carried within the framework of the
Horizon project (\texttt{http://www.projet-horizon.fr}). 
Most of the numerical modeling presented here was done on the Horizon cluster at IAP.
We warmly thank the referee for his/her useful comments 
that improved the quality of the original manuscript.
We warmly thank T.\;Sousbie,  S.\;Prunet, S.\;Rouberol  and M.\;Peeples
for stimulating discussions. We also thank D.\,Munro for
freely distributing his Yorick programming language (available at
\texttt{http://yorick.sourceforge.net/}) which was used during the
course of this work.
\\

\end{document}